\documentclass[a4paper,fleqn,usenatbib]{mnras}

\usepackage{newtxtext,newtxmath}

\usepackage[T1]{fontenc}
\usepackage{ae,aecompl}

\usepackage{graphicx}	
\usepackage{amsmath}	
\usepackage{amssymb}	
\usepackage{todonotes}
\usepackage{bm}         
\usepackage{comment}    
\usepackage{wasysym}     

\newcommand{\dd}[0]{\mathrm{d}}

\newcommand{\rH}[0]{r_{\mathrm{H}}}
\newcommand{\rO}[0]{r_{\mathrm{out}}}
\newcommand{\rB}[0]{r_{\mathrm{B}}}
\newcommand{\lJ}[0]{\ell_{\mathrm{J}}}

\newcommand{\vK}[0]{v_{\mathrm{K}}}

\newcommand{\up}[0]{^{(\mathrm{u})}}
\newcommand{\dn}[0]{^{(\mathrm{d})}}

\newcommand{\edt}[1]{#1}

\graphicspath{ {Pics/} }

\title[Self-gravitating planetary envelopes]{Self-gravitating planetary envelopes and the core-nucleated instability}

\author[W. B\'ethune]{
William B\'ethune$^{1}$\thanks{E-mail: wb288@damtp.cam.ac.uk},
\\
$~^{1}$ Department of Applied Mathematics and Theoretical Physics, University of Cambridge, \\
Centre for Mathematical Sciences, Wilberforce Road, Cambridge CB3 0WA, UK.\\
}

\date{Accepted XXX. Received YYY; in original form ZZZ}

\pubyear{2019}

\begin{document}
\label{firstpage}
\pagerange{\pageref{firstpage}--\pageref{lastpage}}
\maketitle

\begin{abstract}
  Planet formation scenarios can be constrained by the ratio of the gaseous envelope mass relative to the solid core mass in the observed exoplanet populations. One-dimensional calculations find a critical (maximal) core mass for quasi-static envelopes to exist, suggesting that envelopes around more massive cores should collapse due to a `core-nucleated' instability. We study self-gravitating planetary envelopes via hydrodynamic simulations, progressively increasing the dimensionality of the problem. We characterize the core-nucleated instability and its non-linear evolution into runaway gas accretion in one-dimensional spherical envelopes. We show that rotationally-supported envelopes can enter a runaway accretion regime via polar shocks in a two-dimensional axisymmetric model. This picture remains valid for high-mass cores in three dimensions, where the gas gravity mainly adds up to the core gravity and enhances the mass accretion rate of the planet in time. We relate the core-nucleated instability to the absence of equilibrium connecting the planet to its parent disk and discuss its relevance for massive planet formation. 
\end{abstract}


\section{Introduction}

In the common picture of giant planet formation, a gravitational instability (GI) triggers the collapse of gas clumps into protoplanets \citep{podolak93}. The disk instability scenario relies on the GI of massive and cold circumstellar disks \citep{safronov60,toomre64}. This GI can lead to the fragmentation of the disk into substellar companions \citep{boss98,boss2000}, provided an efficient radiative cooling \citep{gammie01,rice03,boothclarke19}. While disks around class 0/I protostars might be massive enough \citep[e.g.,][]{eisner05,ALMA15,liutakami16}, unambiguous signatures of their GI are still looked for \citep{dong16,forgan16,forgan18}. 

Alternatively, planets may grow embedded in the disk \citep{safronov69, goldreichward73}, accumulating dust grains into Earth-sized planetary cores \citep[e.g.,][]{birnstiel16,nimmo18}. As the solid core grows more massive, it attracts more of the surrounding gas in a dense envelope --- its primordial atmosphere \citep{pollack96}. In this core-accretion scenario, planetary envelopes become prone to a GI when massive enough \citep{cameron73}, providing a way to form gas giants within gravitationally stable disks \citep{lissauerpp5}. 

Looking for spherically-symmetric envelope equilibria, \citet{perricameron74} found no solution beyond a \emph{critical core mass}. For a core mass smaller than this critical value, they found two possible solutions for the envelope, and the most massive solution was generally unstable. Since no stable equilibrium can be found beyond the critical mass, the envelope is expected to contract and start accreting gas in a runaway fashion, transforming massive cores into gas giants via the core-nucleated instability.

Subsequent studies aimed at deriving more realistic estimates for the critical core mass. \edt{Different assumptions were made regarding the energy transport through the envelope} \citep{mizuno78,hayashi79,mizuno80,sasaki89}, \edt{the grain opacity and accretion luminosity onto the core} \citep{ikoma01,rafikov06}. \edt{In time dependent models, runaway gas accretion starts when the accretion luminosity can no longer balance the radiative cooling and contraction of the envelope} \citep{pollack96,ikoma2000}. \edt{However, these studies always considered the envelope as one-dimensional and quasi-static.}

\citet{wuchterl2,wuchterl3} presented the first hydrodynamic calculations of one-dimensional gravitating envelopes. At the critical core mass, \citet{wuchterl3} reported a departure from thermal and hydrostatic equilibrium leading to the \emph{ejection} of the envelope. Using three-dimensional hydrodynamic simulations, \citet{ayliffe12} reported the dynamic \emph{collapse} of gravitating envelopes into a more compact equilibrium. What caused the collapse reported by \citet{ayliffe12} and the discrepancies with the results of \citet{wuchterl3} could not be asserted due to the intricate hydrodynamic, chemical and radiative effects involved. 

The aim of this paper is to examine the properties of self-gravitating planetary envelopes in the regime of the core-nucleated instability. We use hydrodynamic simulations in models of increasing dimensionality, keeping simple assumptions for the thermodynamics of the gas. In particular, \emph{we do not} model the runaway cooling, contraction and accretion of radiative envelopes. After examining the response of one-dimensional envelopes at the critical core mass, we consider departures from spherical symmetry by progressively including the rotation and the shear of the flow around the planet. As a number of studies have already characterized three-dimensional non-gravitating envelopes \citep[e.g.,][]{bate03,machida10,fung15}, we focus on the effects induced by the gas gravity on the flow near the core. 

We explicit our model and the methods used throughout this paper in Sect. \ref{sec:method}. We consider one-dimensional envelopes in Sect. \ref{sec:1d}, with hydrostatic calculations followed by hydrodynamic simulations. In Sect. \ref{sec:2d} we consider rotating envelopes within a two-dimensional axisymmetric model. The differential rotation of the circumstellar disk is introduced in Sect. \ref{sec:3d}, where we present three-dimensional simulations of embedded planets in the shearing-sheet approximation. We compare these models against previous studies and discuss their implications in Sect. \ref{sec:discussion}. 

\section{Model and methods} \label{sec:method}

We consider a solid planetary core embedded in a circumstellar disk and massive enough to capture its own atmosphere. If $m_c$ is the mass of the core and $c_s$ the isothermal sound speed of the gas, then the Bondi radius $\rB \equiv G m_c / c_s^2$ is larger than the radius $r_c$ of the core. For simplicity, we consider that the core orbits its star on a circular trajectory, unaffected by the gas drag \citep{weidenschilling77} or other causes of radial migration \citep{kleynelson12}. We consider time intervals of a few tens of orbital periods at most. The properties of the core are fixed and we focus on the dynamics of the gas surrounding it. 

\subsection{Governing equations} \label{sec:equations}

The gas evolves according to the following equations of mass and momentum conservation:
\begin{alignat}{3}
  &\partial_t \rho &&+ \nabla\cdot\left[\rho \bm{v}\right] &&= 0,\label{eqn:consrho}\\
  &\partial_t \left[\rho \bm{v}\right] &&+ \nabla\cdot\left[ \rho \bm{v}\otimes \bm{v} + P \right] &&= - \rho \nabla \Phi - 2 \rho \bm{\Omega}\times \bm{v}, \label{eqn:consrhov}
\end{alignat}
where $\rho$ is the gas density, $\bm{v}$ its velocity and $P$ its pressure. The last term of \eqref{eqn:consrhov} represents the Coriolis acceleration, which we include only when following the core in a frame rotating at the angular frequency $\Omega$. We consider isothermal envelopes in most of this paper, having a pressure $P=\rho c_s^2$ with a single sound speed (temperature) in the entire flow. For a given core mass and disk temperature, the isothermal envelopes are the most massive ones, helping us identify the influence of the gas gravity on the flow. We also consider polytropic envelopes in Sect. \ref{sec:1d}, for which the equation of state is $P = \kappa \rho^\gamma$ and the isothermal limit corresponds to $\gamma=1$. 

We decompose the gravitational potential $\Phi$ as a sum of potentials from the star, the core and the gas. Given the core mass $m_c$, the potential of the core depends on the radius $r$ as $\Phi_c(r) = -G m_c / r$. To avoid more complications, we neglect the gravity of the circumstellar disk: in the limit of $\Phi_c \rightarrow 0$ (no core), we require the potential of the gas to be constant. The potential of the gas must therefore satisfy Poisson's equation
\begin{equation} \label{eqn:poisson}
  \Delta \Phi_g = 4\uppi G \rho',
\end{equation}
in which the source term $\rho'$ is the density deviation from its background value $\rho_{\infty}$. With this source term, an envelope of constant density $\rho_{\infty}$ remains gravitationally stable regardless of its size. If the gas density increases near the core, one can define the Jeans length scale $\lJ^2 \equiv \uppi c_s^2 / G \rho'$ beyond which the gas is unstable to gravito-acoustic perturbations \citep{jeans1902}. 

\subsection{\textsc{Pluto} simulations} \label{sec:plutosetup}

We performed self-gravitating hydrodynamic simulations using a modified version of the \textsc{pluto} 4.0 code \citep{mignone07}. Although the exact numerical setup changes from Sect. \ref{sec:1d} to Sect. \ref{sec:3d}, the integration scheme remains the same for consistency. 

We use \textsc{pluto} to integrate \eqref{eqn:consrho}-\eqref{eqn:consrhov} in time via a finite-volume method and an explicit second-order Runge-Kutta time-stepping. At the volume interfaces, we use a linear reconstruction with the slope limiter of \citet{vanleer79} to estimate the primitive variables $\left(\rho,v\right)$. We then use the approximate Riemann solver of \citet{roe81} to compute the interface fluxes. Where the gas pressure varies by more than a factor $5$ between adjacent cells, we revert to the more diffusive MINMOD slope limiter \citep{roe86} and HLL Riemann solver \citep{van1997relation}. When including the Coriolis acceleration, the momentum equations is evolved in a rotating frame so as to conserve angular momentum \citep{kley98,mignone12}.

We include gravity via its potential $\Phi$. We use the Poisson solver described in Appendix B of \cite{bethune19a} to obtain the potential of the gas satisfying \eqref{eqn:poisson}. We always impose $\partial_r \Phi_g=0$ at the surface of the core, consistently with the absence of gas inside $r\leq r_c$. To avoid a spurious drag arising if the potential lags behind the mass, the Poisson problem is solved at the beginning of every time step. For a spherically symmetric density distribution, the gravitational acceleration $-\partial_r\Phi_g$ could be directly obtained by radial integration of the gas mass. Regardless, we use the same Poisson solver in every dimension for consistency. The numerical error in estimating $\Delta \Phi_g$ is examined in Appendix \ref{app:errors}. 

\subsection{Units and conventions}

We call envelope the region where the core gravity induces a substantial density accumulation $\rho' / \rho_{\infty} \gtrsim 1$. Oppositely, background refers to the conditions in the midplane of the circumstellar disk, near the orbital radius of the core but away from the direct influence of the core. The gravitational constant is set to $G=1$ and we take the radius of the core $r_c$ as distance unit. We use 1D, 2D and 3D in place of one, two and three-dimensional respectively. 

\section{One-dimensional envelopes} \label{sec:1d}

\subsection{1D spherical envelope model}

In this section, we examine the 1D radial structure of spherically-symmetric envelopes. We neglect the orbital motion of the planet around its star, as well as the rotation of the envelope around the core. The only forces involved are the pressure support of the gas against gravity. Were an equilibrium to exist, these two forces would balance each other. We focus on the influence of the gas gravity on such hydrostatic equilibria and their stability. 

Given the background sound speed $c_s$ of the gas, we measure the mass of the core via its Bondi radius $\rB$. We measure the background density $\rho_{\infty}$ relative to the average core density $\rho_c = m_c / \frac{4}{3}\uppi r_c^3$. In the limit $\rho_{\infty}/\rho_c \rightarrow 0$, the gravity of the gas should become negligible compared to the gravity of the core for a finite-sized envelope. 

\subsection{Semi-analytical solutions} \label{sec:semianal}

\subsubsection{Equations}

We design a 1D solver for self-gravitating hydrostatic equilibria. Using a polytropic equation of state $P=\kappa \rho^{\gamma}$, we rewrite the equation of mass \eqref{eqn:consrho} and momentum \eqref{eqn:consrhov} conservation as
\begin{align}
  \frac{\dd m}{\dd r} - 4\uppi r^2 \left[\rho(r)-\rho_{\infty} \right] &= 0,\label{eqn:hs1}\\
  \frac{\dd \log \rho}{\dd \log r} + \frac{\rho^{1-\gamma}}{\gamma\kappa} \frac{m(r)}{r} &= 0, \label{eqn:hs2}
\end{align}
where $m(r)$ is the mass contained inside the sphere of radius $r$ --- core and density deviations. One natural boundary condition is $m\left(r_c\right) = m_c$. We prescribe the second boundary condition at an arbitrary radius $\rO$ and vary the value of $\rO$ so as to simulate the influence of the background density and pressure on the envelope. We impose $\rho(\rO) = \rho_{\infty}$, and set the background sound speed such that $P\left(\rO\right) = \rho\left(\rO\right) c_s^2$ whether $\gamma=1$ (isothermal) or not.

A solution of interest is readily found in the isothermal non-gravitating case. In this limit, \eqref{eqn:hs2} reduces to $\dd\log\rho/\dd\log r + \rB/r = 0$, yielding the non-gravitating solution
\begin{equation} \label{eqn:rhonosg}
\rho_0(r) = \rho_{\infty} \exp\left[\frac{\rB}{r} - \frac{\rB}{\rO} \right]. 
\end{equation}

\subsubsection{Numerical solutions}

We use a Levenberg-Marquardt root finder to solve \eqref{eqn:hs1}-\eqref{eqn:hs2} with the above boundary conditions as a boundary value problem for $\left(m,\log\rho\right)$. The differentiation operators are constructed via a Chebyshev collocation grid on $\log\!\left(r\right)$. With 16 collocation points, the residual error of \eqref{eqn:hs1}-\eqref{eqn:hs2} is less than $10^{-12}$ for smooth solutions; we use 64 collocation points by default. 

\begin{figure}
\begin{center}
\includegraphics[width=1.0\columnwidth]{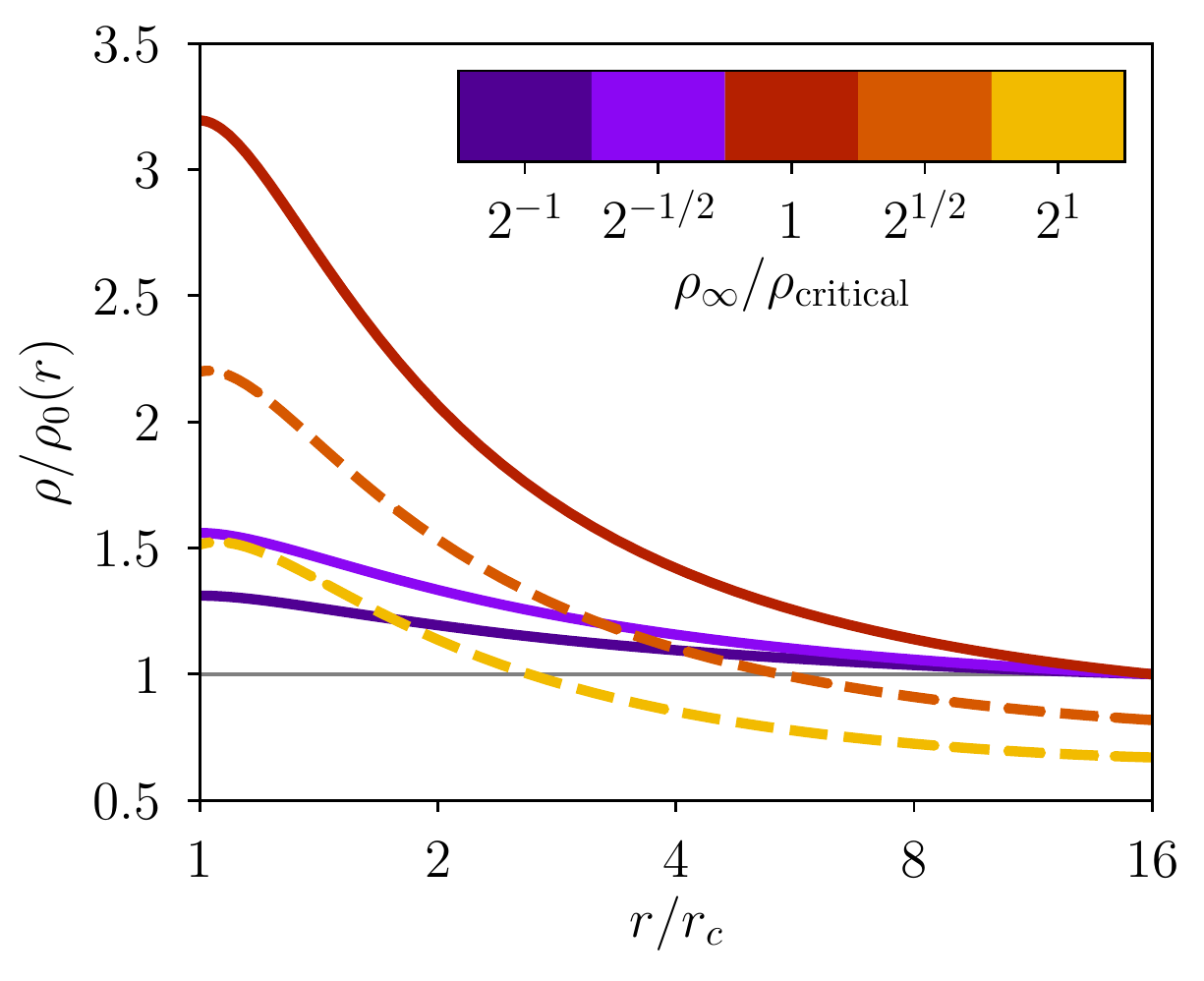}
\caption{Radial density profile normalized by the non-gravitating solution \eqref{eqn:rhonosg} for $\gamma=1$, $\rB/r_c=8$, $\rO/r_c=16$, and varying the background density $\rho_{\infty}$. The dashed curves are obtained when trying to impose $\rho_{\infty}>\rho_{\mathrm{critical}}$; they satisfy \eqref{eqn:hs1}-\eqref{eqn:hs2} but not the required outer boundary condition. \label{fig:hsprofs}}
\end{center}
\end{figure}

Fig. \ref{fig:hsprofs} shows the results of the root finder for $\gamma=1$, $\rO/r_c=16$, $\rB/r_c=8$, and different values of the background density $\rho_{\infty}$. In the non-gravitating limit $\rho_{\infty}\rightarrow 0$ (darker, violet lines), the solution converges to \eqref{eqn:rhonosg}. As $\rho_{\infty}$ increases, the gas gravity becomes significant and adds up to the core gravity. To maintain a hydrostatic equilibrium, the pressure (density) profile becomes progressively steeper. Eventually, we reach a critical value $\rho_{\mathrm{critical}}$ beyond which there is no \emph{valid solution} of \eqref{eqn:hs1}-\eqref{eqn:hs2} satisfying $\rho(\rO)=\rho_{\infty}$ (lighter, orange curves). We identify this transition by monitoring the error $\rho(\rO)-\rho_{\infty}$, which suddenly jumps at the threshold $\rho_{\infty} = \rho_{\mathrm{critical}}$. 

Although the outer boundary condition is not satisfied anymore, the density profiles in the regime $\rho_{\infty}>\rho_{\mathrm{critical}}$ are still solution to \eqref{eqn:hs1}-\eqref{eqn:hs2} to per cent accuracy over the radial domain. The critical density marks the transition to the second solution branch identified by \cite{perricameron74}, albeit with a different parametrization. We examine the linear stability of these solutions in Appendix \ref{app:linstab}, and now delimit the domain of existence of solutions satisfying $\rho\left(\rO\right)=\rho_{\infty}$ in our parameter space. 

\subsection{Critical envelopes} \label{sec:critmass}

\begin{figure}
\begin{center}
\includegraphics[width=1.0\columnwidth]{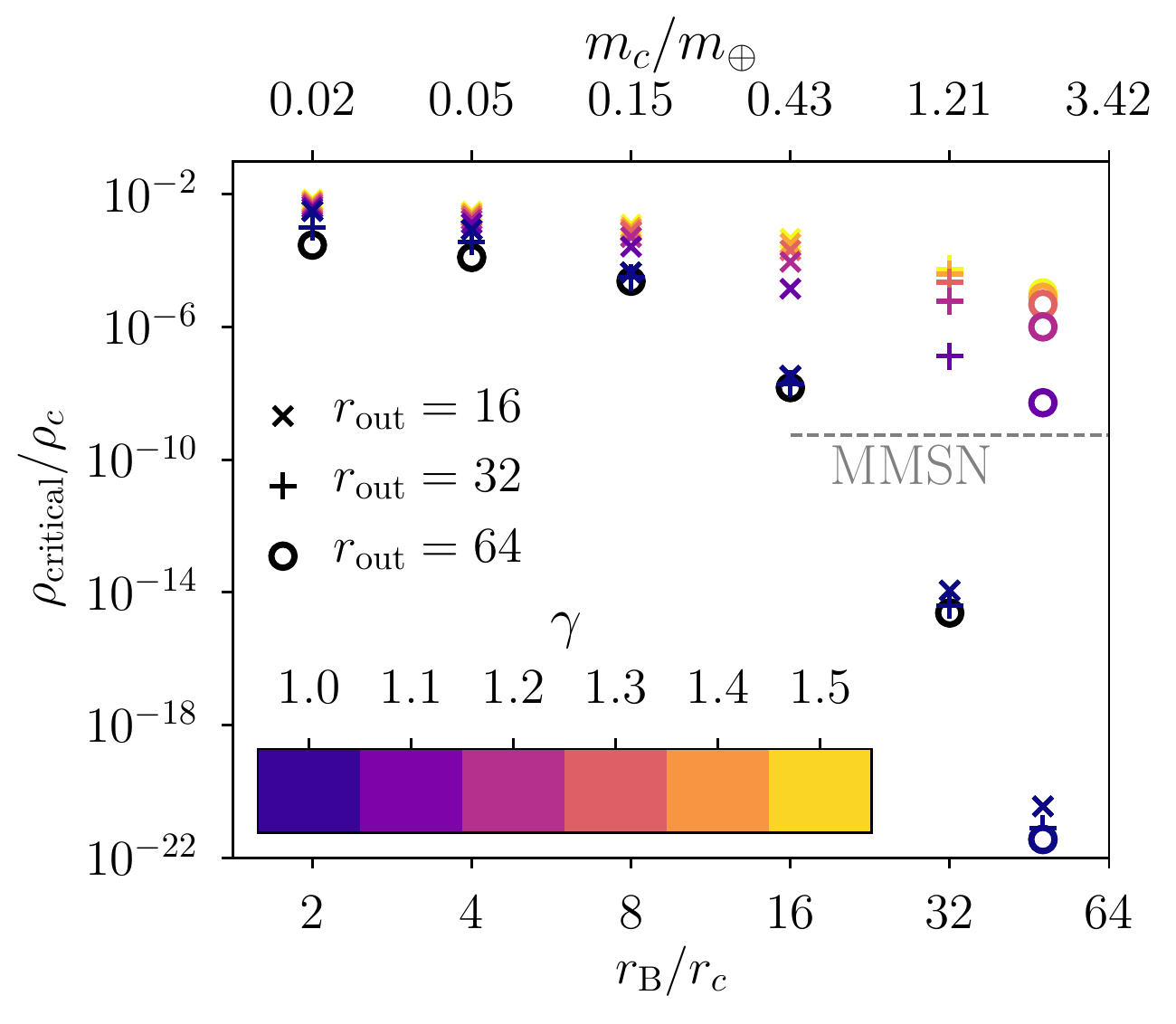}
\caption{Critical background density relative to the core density for different values of the Bondi radius $r_{\mathrm{B}}$ (\emph{abscissa}), outer radius $\rO/r_c$ (\emph{markers, see legend}) and polytropic exponent $\gamma$ (\emph{colors scale}). Valid solutions are found below the markers, i.e. for smaller $\rho_{\infty}/\rho_c$. The top axis indicates the corresponding core mass in Earth mass units, assuming that the core has the same density as the Earth and is located at $1\mathrm{au}$ around a solar-mass star in a disk with aspect ratio $h/r=0.05$. The dashed horizontal line marks the corresponding MMSN midplane density $\rho\approx 3\times10^{-9} \mathrm{g}\,\mathrm{cm}^{-3}$. \label{fig:ssk_stablim}}
\end{center}
\end{figure}

We solve \eqref{eqn:hs1}-\eqref{eqn:hs2} for various input parameters and track the threshold value $\rho_{\mathrm{critical}}$. We prescribe the outer radius $\rO/r_c\in\left[16,64\right]$ to represent the sphere of influence of a massive embedded core, i.e. roughly one pressure scale of the disk in radius \cite[see Sect. \ref{sec:3dflow} and][]{bethune19b}. We also prescribe the polytropic exponent $\gamma$ while maintaining\footnote{For static equilibria, one can arbitrarily choose $c_s$ and adjust the core mass according to $m_c = c_s^2 \rB / G$.} the background temperature $P(\rO)=\rho_{\infty} c_s^2$, and the mass of the core via its Bondi radius $\rB$. We then increase the background density $\rho_{\infty}$ until the outer boundary condition can no longer be satisfied. The corresponding $\rho_{\mathrm{critical}}$ are marked on Fig. \ref{fig:ssk_stablim} relative to the core density $\rho_c$. 

The ratios of $\rho_{\mathrm{critical}}/\rho_c$ marked on Fig. \ref{fig:ssk_stablim} delimit the region of existence of valid solutions from above. With $r_c$ as our distance unit, the core mass is directly $m_c \sim \left(4/3\right)\uppi \rho_c$. For a given core mass, valid equilibria require a background density $\rho_{\infty}$ smaller than $\rho_{\mathrm{critical}}$. Reciprocally, solutions satisfying $\rho_{\infty}=\rho_{\mathrm{critical}}$ can only be found for cores more massive than indicated. 

We note three trends on Fig. \ref{fig:ssk_stablim}. First, when the mass of the core increases (from left to right), the critical gas-to-core density ratio decreases. Second, increasing the polytropic exponent $\gamma$ from $1$ to $3/2$ (lighter markers) allows equilibria at larger background densities. Third, the threshold $\rho_{\mathrm{critical}}/\rho_c$ only slightly decreases with the location of the outer boundary condition $\rO$ (marker symbols). 

The thresholds obtained by this method agree with those of \citet{sasaki89} in the appropriate regime. The critical density drops by orders of magnitude as soon as $\rB/r_c\gtrsim 8$. It falls below the Minimum Mass Solar Nebula \citep[MMSN,][]{hayashi81} midplane density for isothermal envelopes around cores of a few Earth masses at $1\,\mathrm{au}$. However, valid equilibria with $\gamma=7/5$ still exist for much more massive cores $\gtrsim 10^3 m_{\oplus}$. Whether the critical density threshold is realistically accessible for more sophisticated thermodynamic structures is outside the scope of this paper.

Having delimited the range of parameters for which valid hydrostatic equilibria exist, we proceed to examine the dynamical reaction of the envelope when the control parameters vary continuously across this limit. 

\subsection{Direct numerical simulations}

\subsubsection{Numerical setup} \label{sec:1dsetup}

We use the \textsc{pluto} code as described in Sect. \ref{sec:plutosetup} to evolve the density $\rho$ and radial velocity $v_r$ in time for an isothermal gas in 1D spherical geometry. We mesh the radial interval $r/r_c \in \left[1,16\right]$ with 512 logarithmically spaced grid cells. We prevent mass and momentum fluxes through the surface of the core via $\left(\rho,v_r,\Phi_c\right)(r_c-\epsilon) = \left(+\rho,-v_r,+\Phi_c\right)(r_c+\epsilon)$. At the outer radial boundary, we impose a constant density $\rho_{\infty}$ and allow the gas to flow in by a linear extrapolation of the radial velocity. The outer boundary condition $\Phi_g(\rO)=0$ sets a reference value for the gravitational potential of the gas. We initialize the computational domain with a flat density $\rho=\rho_{\mathrm{\infty}}$ and zero velocity. We take the sound speed as velocity unit ($c_s=1$). The mass of the core is progressively increased from zero to its nominal value, so that the envelope mass builds up in quasi-static equilibrium at every instant. 

\subsubsection{Non-gravitating limit} \label{sec:dns1dnosg}

\begin{figure}
\begin{center}
\includegraphics[width=1.0\columnwidth]{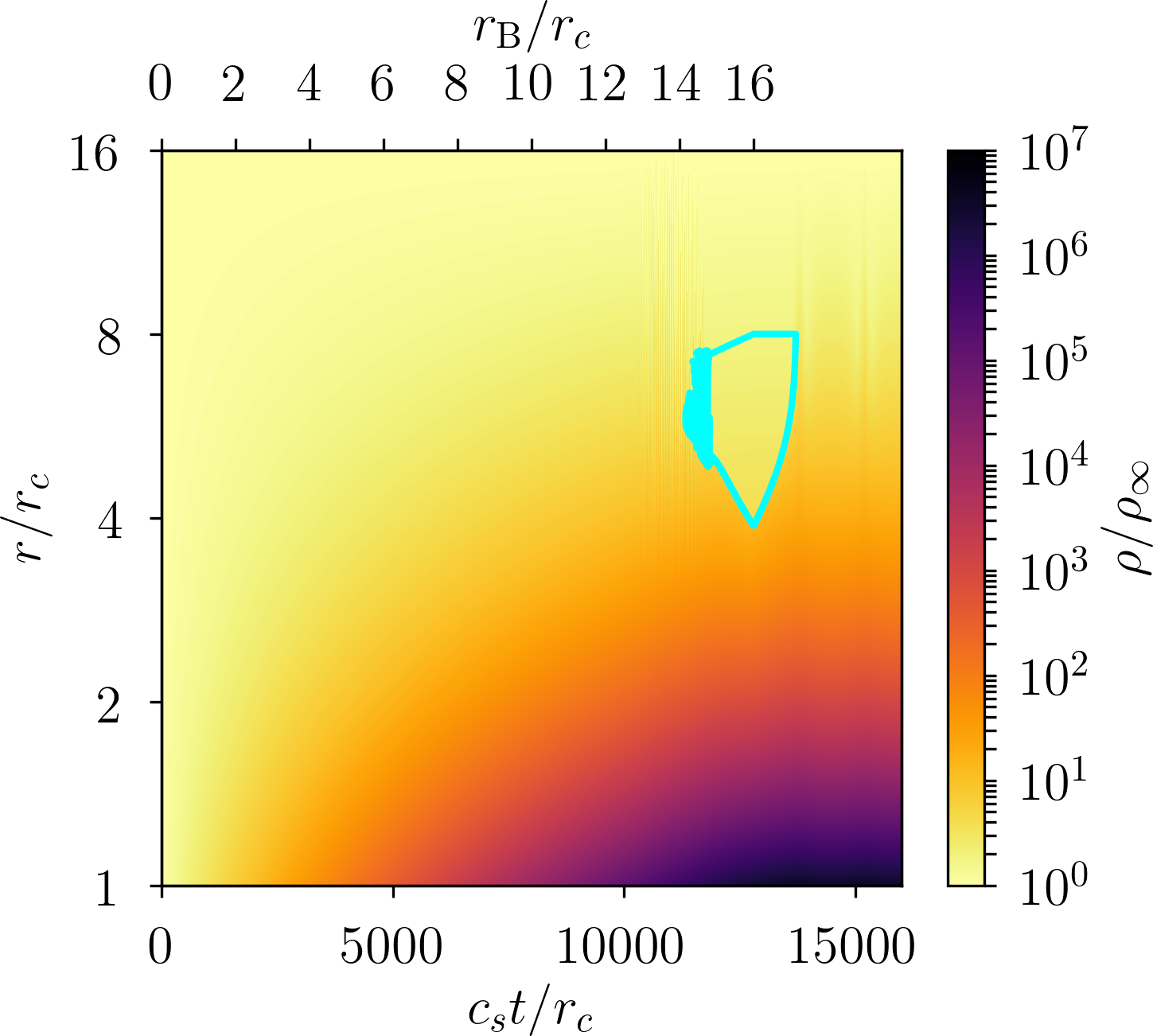}
\caption{Space-time distribution of the density $\rho/\rho_{\infty}$ in a non-gravitating envelope. The core mass increases from zero to $\rB=16\,r_c$ over the first $t_{\mathrm{B}}=12800 r_c/c_s$ (\emph{top axis}) after which it is constant. The cyan contour marks the sonic surface $v_r=-c_s$, corresponding to a smooth transsonic point (upper part) and a shock (lower part). \label{fig:rhortnosg}}
\end{center}
\end{figure}

When neglecting the gas gravity, the density profile should converge toward \eqref{eqn:rhonosg} given an outer density $\rho_{\infty}$ and a Bondi radius $\rB$. We increase the Bondi radius of the core linearly in time from zero up to $16\,r_c$ over a time interval of $t_{\mathrm{B}} = 12800\,r_c/c_s$, after which it remains equal to $16\,r_c$. The resulting density distribution is represented on Fig. \ref{fig:rhortnosg} as a function of radius and time. 

As the Bondi radius of the core increases (from left to right), an inflow of gas through the outer radial boundary allows the density to increase inside the domain. At $c_s t / r_c \approx 12000$ and $r/r_c \approx 8$, the inflow becomes supersonic (cyan contour) and shocks on the inner parts of the envelope. The shock front propagates inward until $t_{\mathrm{B}}$, i.e., while the mass of the core is still increasing. For $t>t_{\mathrm{B}}$, the shock front propagates outward until the two sonic points merge. After this instant, the envelope sustains acoustic oscillations but matches the analytical solution \eqref{eqn:rhonosg} to $5$ per cent accuracy upon time-averaging. 

The shock appearing in this simulation is a consequence of the inflow velocity exceeding the sound speed. It implies that the envelope is not in quasi-static equilibrium as intended, certainly because the Bondi radius of the core initially increases too fast. However, the envelope is able to reach a stable equilibrium after the core mass stops increasing. 

\subsubsection{Self-gravitating envelope}

We repeat the same simulation as above, but now including the gravity of the gas. For a Bondi radius $\rB=16\,r_c$, the critical density is $\rho_{\mathrm{critical}}/\rho_c \approx 3.346 \times 10^{-8}$. We impose a slightly larger density $\rho_{\infty}\approx 3.356 \times 10^{-8} \rho_c$. When the Bondi radius of the core reaches $16\,r_c$, no hydrostatic equilibrium should be able to satisfy the outer boundary condition anymore.

\begin{figure}
\begin{center}
\includegraphics[width=1.0\columnwidth]{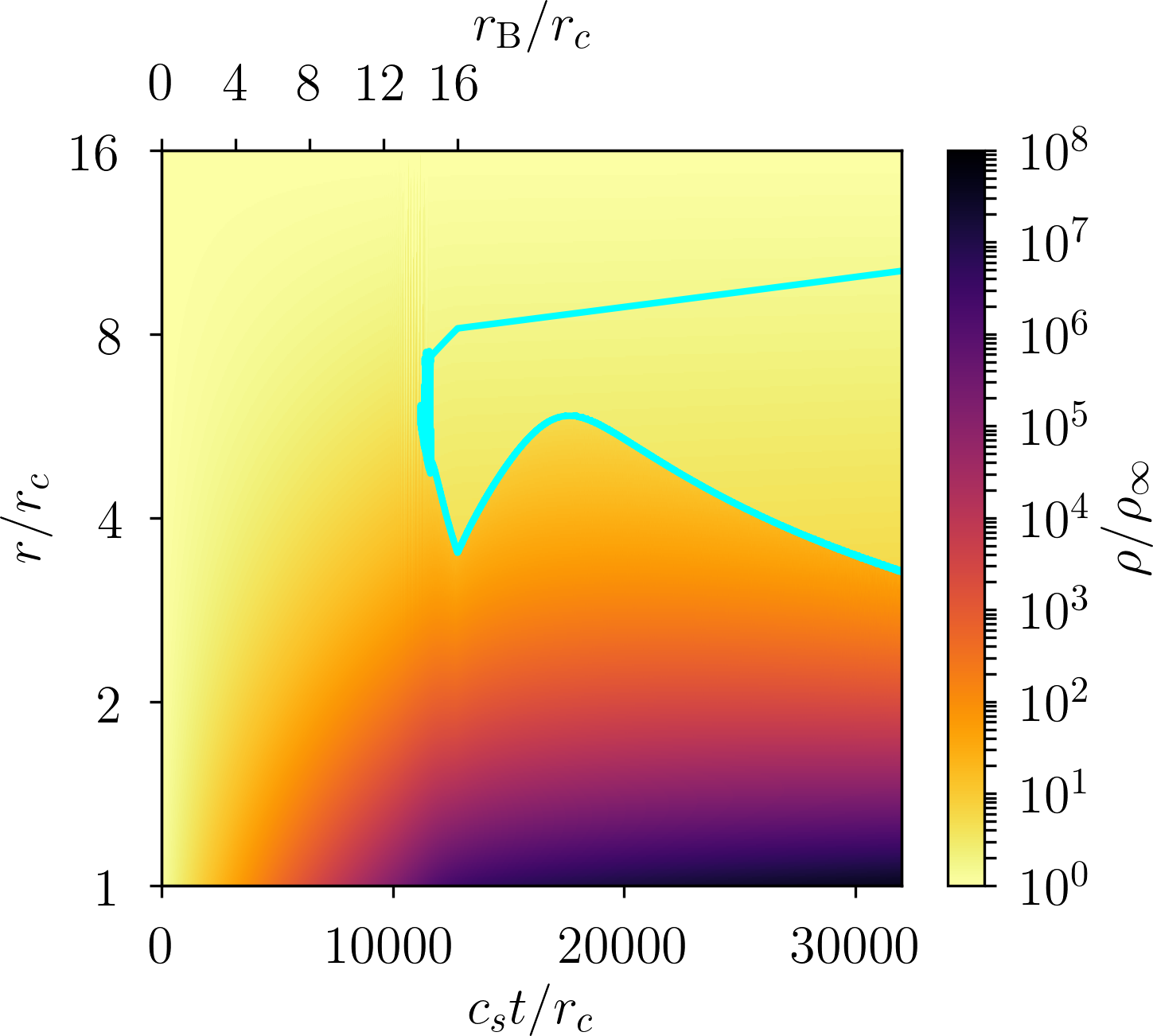}
\caption{Same as Fig. \ref{fig:rhortnosg} but including the gas gravity, with a background density larger than its critical value. The inflow remains transsonic after the Bondi radius has reached $\rB=16\,r_c$; the envelope is then delimited by an inward moving shock front (lower part of the cyan contour). \label{fig:rhortwithsg}}
\end{center}
\end{figure}

The evolution of the self-gravitating density distribution is shown on Fig. \ref{fig:rhortwithsg}. As previously, the inflow becomes supersonic before the Bondi radius of the core reaches $16\,r_c$. After $t_{\mathrm{B}}=12800\,r_c/c_s$, the shock moves outward for about $5\times 10^3\,r_c/c_s$ before starting to move toward the core. The smooth sonic point (upper part of the cyan line) keeps moving outward, meaning that the inflow becomes progressively faster at a given radius. As the two sonic points drift away from one another, the core and its envelope always drive a transsonic inflow. 

The key difference with the non-gravitating case is that the envelope does not converge to a steady state when $\rho_{\infty}/\rho_c$ exceeds the threshold delimited on Fig. \ref{fig:ssk_stablim}. When crossing this threshold, the outer parts of the envelope collapse in near free-fall. As soon as the infalling gas becomes supersonic, it has to shock on the inner parts of the envelope, dissipating momentum and allowing the accumulation of mass.

We verified via simulations at lower core mass (lower $\dd r_{\mathrm{B}}/\dd t$ while the core mass is initially increased) and larger $\rho_{\infty}$ that the collapse precisely occurs at the critical values predicted in Sect. \ref{sec:critmass}. However, in this regime the inflow shocks closer to the core and the shocked envelope becomes spatially under-resolved. We verified that a collapse also happens in adiabatic envelopes when crossing the threshold marked on Fig. \ref{fig:ssk_stablim}. In this case, adiabatic heating and momentum dissipation at the shock lead to higher pressures in the contracting gas. For $\gamma=3/2$, $\rO/r_c=16$ and $\rB/r_c=16$, the smooth sonic point settles at the outer radius $\rO$ and the shock propagates \emph{outward} until the space between the shock and $\rO$ is under-resolved. Because these cases are sensitive to the finite extent and resolution of our computational domain, we discuss what governs the dynamics of the envelope in the next section. 

\subsection{Discussion of 1D models} \label{sec:1ddiscussion}

\subsubsection{The core-nucleated instability}

By progressively increasing the mass of the core at a fixed $\rho_{\infty}$, we have followed the `stable' solution branch for the envelope up to the critical core mass. Beyond this point, the equilibria from the second solution branch (as drawn on Fig. \ref{fig:hsprofs} for $\rho_{\infty}>\rho_{\mathrm{critical}}$) should be linearly unstable \citep{perricameron74,mizuno78,wuchterl1}. Due to the sudden collapse of the outer envelope at the critical mass, these equilibria seem inaccessible unless taking them as initial condition \citep[as done by][]{wuchterl3}. We will therefore focus on the non-linear dynamics of the envelope when it crosses the critical mass from below. 

It is possible to interpret the trends of Fig. \ref{fig:ssk_stablim} in the non-gravitating limit, assuming that the threshold mainly depends on the envelope mass relative to the core mass \citep{sasaki89}. First, the mass of the envelope increases faster than the mass of the core, so gas gravity effects appear at lower background densities $\rho_{\infty}$ when the mass of the core increases. Second, if the gas pressure varies as $\rho^{\gamma}$, then one can satisfy $\partial_r P = -\rho\partial_r\Phi_c$ with a shallower density profile when increasing $\gamma$. To reach the same envelope mass, the background density must then be larger. Third, the envelope mass increases with $\rO$, so it becomes comparable to the core mass at a lower $\rho_{\mathrm{critical}}$ when $\rO$ increases. 

As we show on Fig. \ref{fig:ssk_stablim} and in Appendix \ref{app:linstab}, the absence of global equilibria beyond a critical mass is independent of the gas thermodynamics, which only affect the value of this critical mass. Unlike the instability of a homogeneous gas ball \citep{ebert55,bonnor56}, the envelope collapse can spontaneously stop even though $\gamma<4/3$ (see Fig. \ref{fig:rhortwithsg}). As apparent from the slow propagation of the shock front on Fig. \ref{fig:rhortwithsg}, the shocked envelope maintains a nearly hydrostatic equilibrium. Using the method described in Sect. \ref{sec:semianal}, we verified that hydrostatic equilibria can indeed be found between the core and the shock, given the post-shock density as an outer boundary condition. In this sense, planetary cores above the critical mass can still support a hydrostatic envelope. We explain below how the extent of this envelope is determined by the ambient conditions. 

\subsubsection{Long-term evolution} \label{sec:1dlongterm}

In 1D, if the infalling gas becomes supersonic, then it has to shock before reaching the surface of the core (where $v_r=0$). The conditions just upstream of the shock are controled by the ambient (outer boundary) conditions on $\left(\rho,v_r\right)$ and by the planet mass. Given the ambient conditions, it is possible to predict the dynamics of 1D envelopes to some extent. 

Let $\zeta(t)$ denote the radius of the shock front, $u_r$ the gas velocity in the frame of the shock, and the exponents $(\mathrm{u})$ and $(\mathrm{d})$ identify the upstream and downstream regions respectively. The velocity of the shock is obtained by changing frame: $\dd\zeta/\dd t = v_r\dn - u_r\dn$. If the post-shock envelope was exactly hydrostatic ($v_r\dn=0$), then $\dd\zeta/\dd t>0$ and the shock would propagate outward. The opposite orientation $\dd\zeta/\dd t<0$ on Fig. \ref{fig:rhortwithsg} reveals that the post-shock envelope is contracting ($v_r\dn<0$). This contraction is due in part to the inward momentum flux, most of it being converted into pressure at the shock. Simultaneously, the accumulation of mass causes the gas potential $\Phi_g$ to deepen over time, so the envelope contracts to support its own increasing gravity. 

The momentum flux is relevant to locate the shock front. The velocity drop at the shock --- $u_r\dn/c_s = c_s/u_r\up$ in the isothermal case --- leads to a large drop in ram pressure $\rho u_r^2$. The shock ultimately settles where the downstream thermal pressure balances the upstream ram pressure: $\rho\dn c_s^2 \simeq \rho\up u_r\up{^2}$. Since the shocked envelope is nearly hydrostatic, $\rho\dn$ decreases radially, so a larger upstream ram pressure would push the shock closer to the core. This is expected if the background density or the accumulated envelope mass increases (as on Fig. \ref{fig:rhortwithsg}). 

If the ambient density were to decrease, then the shocked shell would expand until the momentum balance condition is satisfied again. If the shock radius $\zeta$ extends further than the gravitational radius $G m\left(\zeta\right) / c_s^2$ or the Hill radius of the planet, then the outer parts of the envelope could escape the core by evaporation or gravitational tides. According to this 1D model, a planet may therefore experience a phase of rapid gas accretion only to lose its outer envelope later during the dispersal of the protoplanetary disk. 


\section{Two-dimensional envelopes} \label{sec:2d}

The previous 1D model omits the angular momentum of the background flow with respect to the core. By conservation of angular momentum, the gas should spin faster as it approaches the core. The centrifugal acceleration can then provide a substantial support against gravity, allowing for much less massive envelopes. In this section, our main goal is to test whether the core-nucleated instability --- and the ensuing accretion phase --- can also affect rotationally-supported envelopes.

\subsection{2D axisymmetric model} \label{sec:2dmodel}

We consider a planetary core orbiting its star at the angular frequency $\Omega$ about the $z$ axis of the disk. We adopt a frame centered on the core and rotating along its orbit at the angular frequency $\Omega$; in this frame, the background flow is steady in time. To make things simpler, we neglect the vertical stratification and the differential rotation of the disk. In the absence of the core, the gas density should be constant and the velocity should be zero in this frame, so rotation only manifests itself through the Coriolis acceleration in \eqref{eqn:consrhov}. If radial motions are brought about by the core, then angular momentum conservation will generate a toroidal velocity in this frame. 

Let $\left(r,\theta,\varphi\right)$ denote spherical coordinates centered on the core, with $\theta=0$ along the rotation axis $z$. Assuming that the flow is axisymmetric ($\partial_{\varphi}=0$), we use \textsc{pluto} to integrate \eqref{eqn:consrho}-\eqref{eqn:consrhov} for $\left(\rho,v_r,v_{\theta},v_{\varphi}\right)$ in the $\left(r,\theta\right)$ poloidal plane. 

The interval in polar angle $\theta\in \left[0,\uppi\right]$ is uniformly meshed with 256 grid cells. The radial interval $r/r_c\in\left[1,32\right]$ is meshed with 256 logarithmically spaced cells. The radial boundary conditions are the same as in 1D (see Sect. \ref{sec:1dsetup}), with the addition of $v_{\theta}=v_{\varphi}=0$. Inside the computational domain, we homogenize the density in the innermost grid shell at every timestep, conserving the total mass in the shell. This operation is intended to prevent unresolved mass concentrations when including the gas gravity. About the polar axis $\theta=0$, we impose 
\begin{equation*}
\left(\rho,v_r,v_{\theta},v_{\varphi}\right)(r,-\vartheta) = \left(+\rho,+v_r,-v_{\theta},-v_{\varphi}\right)(r,+\vartheta),
\end{equation*}
and similarly about $\theta=\uppi\pm\vartheta$. We initialize the domain with $\rho=\rho_{\infty}$ and $\bm{v}=0$, and we increase the mass of the core up to its nominal value linearly over $\Omega t/2\uppi = 2$ orbital times. The gas entering the computational domain carries a specific angular momentum depending only on its initial latitude and on the outer radius $\rO$.

\subsection{Non-gravitating limit} \label{sec:dns2dnosg}

We start by neglecting the gravity of the gas. We take $\rho_{\infty}=c_s=1$ and prescribe the angular frequency $\Omega r_c / c_s = 1/32$ of the core around the star. If the vertical stratification of the disk was accounted for, the density would vary over a pressure scale height $h\equiv c_s/\Omega$. The ratio $\Omega r_c / c_s$ would then measure the size of the core relative to the stratification scale of the disk. Typical values of this ratio for super-Earths can be found in section 2.1 of \cite{bethune19a}. A ratio of $h/r_c = 32$ is arguably reasonable for massive planets at small orbital separations, but under-estimated otherwise. Larger values of $h/r_c$ would place stronger constraints on the explicit integration time steps and therefore be computationally more demanding. We consider two cases with different core masses: $r_{\mathrm{B}}/r_c=8$ and $16$. 

\subsubsection{Flow structure} \label{sec:dns2dflow}

\begin{figure}
\begin{center}
\includegraphics[width=1.0\columnwidth]{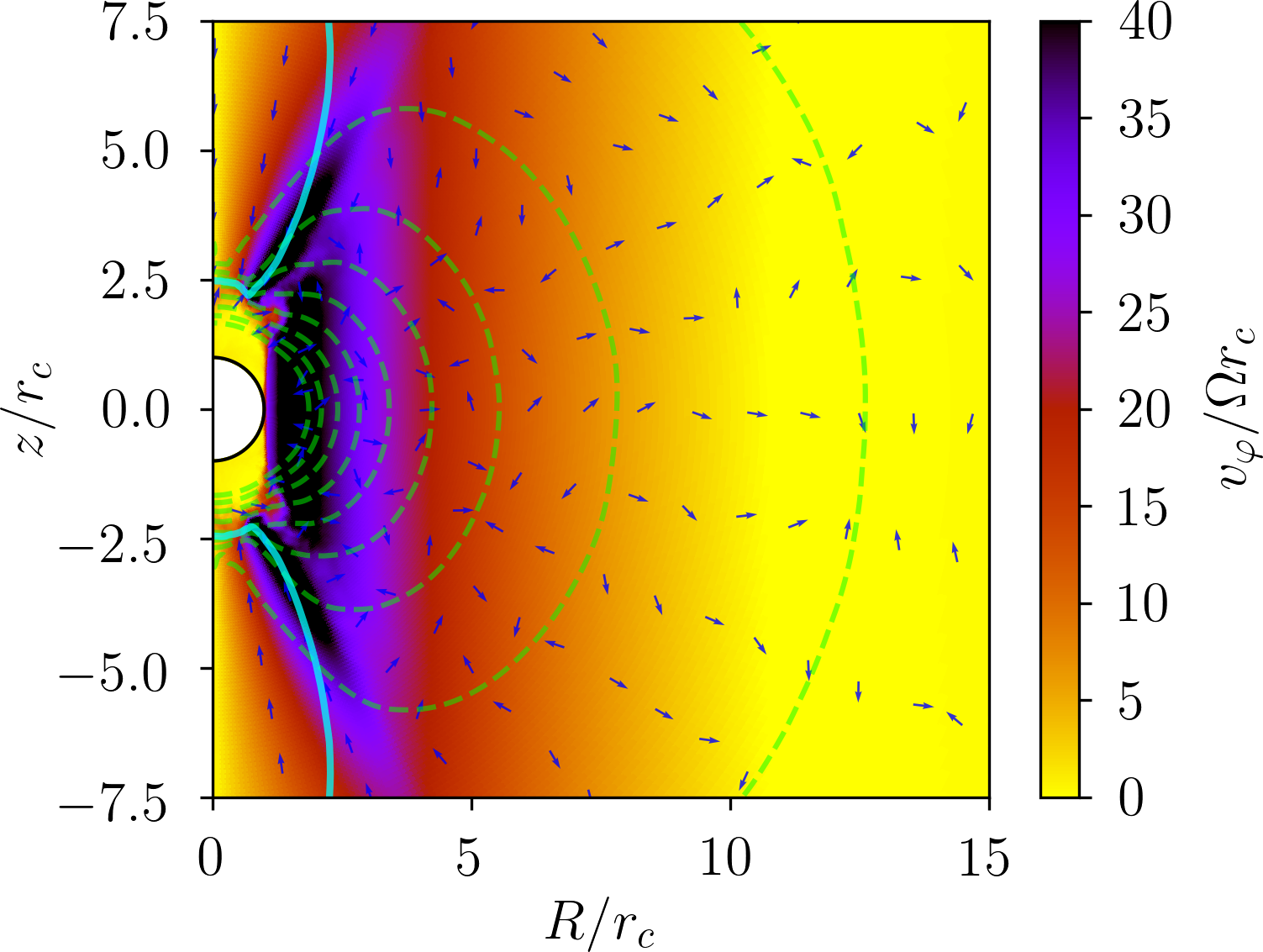}
\caption{Toroidal velocity (\emph{color map}), poloidal mass flux (\emph{blue arrows, orientation only}), mass density (\emph{dashed green contour lines}) and sonic surface for the poloidal velocity (\emph{solid cyan contours}) after time-averaging over $20\Omega^{-1}$ in the simulation with $\rB/r_c=16$. Note the non-rotating shell directly on top of the core. \label{fig:2dvyrhovp}}
\end{center}
\end{figure}

Fig. \ref{fig:2dvyrhovp} shows the time-averaged flow in the non-gravitating simulation with $\rB/r_c=16$. In the midplane, the toroidal velocity increases inward, as expected from angular momentum conservation \citep{miki82,ormel1}. \edt{To quantify the level of rotational support, one can convert to Keplerian velocity units}
\begin{equation} \label{eqn:tokepler}
  \frac{v_{\varphi}}{\vK} = \frac{\Omega r_c}{\sqrt{c_s^2\rB/R}}\frac{v_{\varphi}}{\Omega r_c}  = \frac{1}{32} \sqrt{\frac{R}{\rB}} \frac{v_{\varphi}}{\Omega r_c}.
\end{equation}
The toroidal velocity increases from $0.2\vK$ at $R=6 r_c$ to its maximum $0.4\vK$ at $R=2r_c$in the midplane. \edt{Since the centrifugal acceleration scales as $v_{\varphi}^2$}, this corresponds to roughly $4$ to $16$ per cent of rotational support against gravity. 

Close to the polar axis, the gas comes with essentially no angular momentum, so only the pressure gradient can balance gravity. However, the envelope does not settle in a static 1D equilibrium. Instead, the gas circulates from high latitudes down to the core and away from the core in the midplane. The inflow become supersonic at $\vert z\vert \approx 7r_c$, and subsonic again through a shock at $\vert z \vert \simeq 2.5r_c$. Averaging the turbulent fluctuations out, the radius of the shock increases by less than $0.2 r_c$ over $3000$ sound crossing times. With respect to the core, the shock front is therefore stationary over the time scales considered, and it dissipates the momentum of the infalling gas. Downstream of the shocks, the gas remains sitting on top of the core with essentially no momentum. 

Whether an accretion shock forms relies on the inflow becoming supersonic. Otherwise, the envelope can be recycled by the poloidal circulation with no net mass accretion onto the core \citep{ormel2,bethune19b}. We now characterize the mass flux through the envelope for different core masses. 

\subsubsection{Gas accretion and recycling}

\begin{figure}
\begin{center}
\includegraphics[width=1.0\columnwidth]{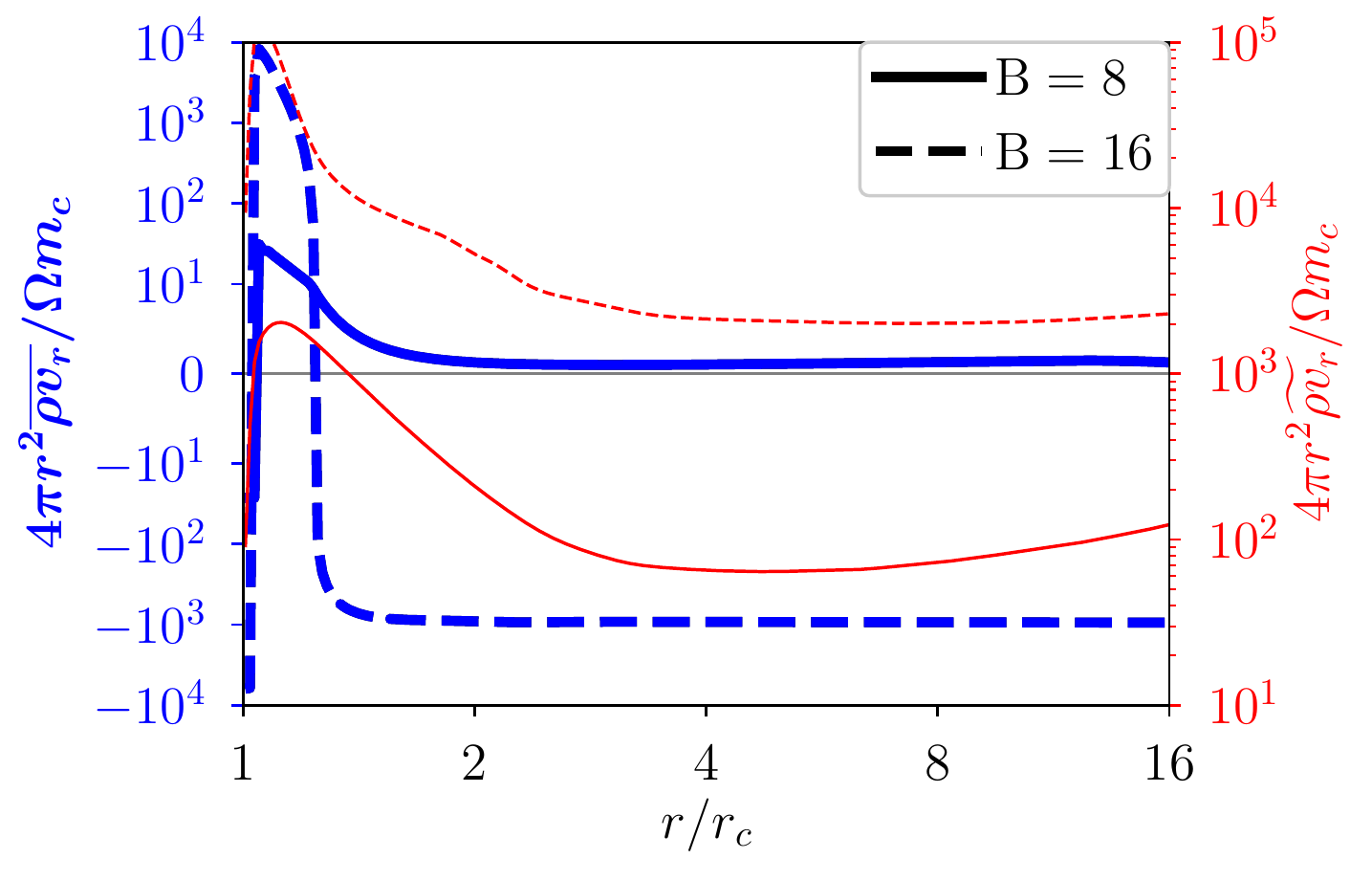}
\caption{Radial profiles of the net mass flux $\overline{\rho v_r}$ (\emph{thick blue, left axis}) and recycling mass flux $\widetilde{\rho v_r}$ (\emph{thin red, right axis}) averaged over $50\Omega^{-1}$ in non-gravitating 2D simulations with $\rB/r_c=8$ (\emph{solid}) and $16$ (\emph{dashed}). \label{fig:2dmvrdvr}}
\end{center}
\end{figure}

Let $\overline{X}(r)$ denote the average of $X$ over the sphere of radius $r$. From the net mass flux $\overline{\rho v_r}$, we define the recycling mass flux $\widetilde{\rho v_r}$ as the standard deviation $\widetilde{\rho v_r}^2 \equiv \overline{\rho^2 v_r^2} - \overline{\rho v_r}^2$. The recycling flux is zero when the flow is spherically symmetric, but non-zero if the flow features some degree of circulation. 

We draw the radial profiles of the net and recycling mass fluxes in the $\rB/r_c=8$ and $16$ cases on Fig. \ref{fig:2dmvrdvr}. The recycling flux $\widetilde{\rho v_r}$ is non-zero in both cases, so both envelopes support a poloidal circulation. The recycling mass flux is maximal closer to the core, and it increases by a factor 10--20 when $\rB/r_c$ increases from $8$ to $16$. Because of the shell averaging, the recycling flux $\widetilde{\rho v_r}$ in the $\rB/r_c=16$ case is non-zero all the way down to the core, although the shocked gas on top of the core does not seem to be efficiently mixed with its surrounding on Fig. \ref{fig:2dvyrhovp}.

Regarding the net accretion flux $\overline{\rho v_r}$, we obtain two qualitatively different behaviors depending on the core mass. For $\rB/r_c=16$, the accretion rate $4\uppi r^2 \overline{\rho v_r} \approx - 10^{3} \Omega m_c$ is constant throughout the computational domain. The positive mass flux measured below $1.25r_c$ is most likely a representation artifact due to the different variables used by \textsc{pluto} and in the present analysis\footnote{\textsc{Pluto} evolves the conservative variables $\left(\rho,\rho \bm{v}\right)$ after reconstruction of the primitive variables $\left(\rho,\bm{v} \right)$ at the cell interfaces and using the fluxes of the Riemann problem. For our analysis, we estimate the mass flux after time-averaging the cell centered primitive variables.}. We verified that the accretion rate $\partial_t m(r)$ matches the integrated flux $-4\uppi r^2 \overline{\rho v_r}(r)$ to better than $10^{-2}$ relative accuracy, so there is no measurable mass flux through the core boundary. 

For $\rB/r_c=8$, the net mass flux $\overline{\rho v_r}$ drawn on Fig. \ref{fig:2dmvrdvr} is compatible with zero through the envelope. The mass cumulated inside the Bondi sphere oscillates by $5$ per cent about its equilibrium value, with no net increase over $900$ sound crossing times of the Bondi sphere. The absence of mass accretion is related to the absence of dissipative processes, and specifically the absence of accretion shocks. The poloidal velocity is indeed subsonic everywhere in this run, recycling the envelope without accumulating mass on top of the core \citep{bethune19b}. 

\subsection{Self-gravitating axisymmetric envelopes} \label{sec:2dsg}

We now include the gravity of the gas in addition to rotation. To facilitate comparisons with later 3D results, we adopt $\Omega=1$ while keeping $\Omega r_c / c_s=1/32$. For comparison purposes again, we pretend that the planet is embedded in a Keplerian shear flow, and that the disk is stratified over a pressure scale height $h$. The background density can then be prescribed in terms of the Toomre parameter
\begin{equation} \label{eqn:toomre}
Q \equiv \frac{\Omega c_s}{\uppi G \Sigma} \sim \frac{1}{\uppi \sqrt{2\uppi} \rho_{\infty}}
\end{equation}
with our choice of units. We consider five different combinations of $r_{\mathrm{B}}/r_c$ and $Q$, as listed in \autoref{tab:2dsg}.

\begin{table}
\begin{center}
\caption{Two-dimensional self-gravitating simulations: label, Bondi radius $r_{\mathrm{B}}/r_c$, Toomre parameter $Q$ as defined by \eqref{eqn:toomre}, background density $\rho_{\infty}$ relative to the core density $\rho_c$, critical background density for the equivalent 1D setup, and existence of a correspondig 1D hydrostatic equilibrium. \label{tab:2dsg}}
\begin{tabular}{lccccc}
Label & $\rB/r_c$ & $Q$ & $\rho_{\infty}/\rho_c$ & $\rho_{\mathrm{critical}}/\rho_c$ & 1D static\\
\hline
\verb!2B8Q0!  &  $8$ & $10^{0}$   & $6.49\times 10^{-5}$ & $2.95\times 10^{-5}$ & no\\
\verb!2B8Q05! &  $8$ & $10^{0.5}$ & $2.05\times 10^{-5}$ & $2.95\times 10^{-5}$ & yes\\
\verb!2B8Q1!  &  $8$ & $10^{1}$   & $6.49\times 10^{-6}$ & $2.95\times 10^{-5}$ & yes\\
\verb!2B16Q1! & $16$ & $10^{1}$   & $3.24\times 10^{-6}$ & $1.10\times 10^{-8}$ & no\\
\verb!2B16Q2! & $16$ & $10^{2}$   & $3.24\times 10^{-7}$ & $1.10\times 10^{-8}$ & no
\end{tabular}
\end{center}
\end{table}

\subsubsection{Envelope mass} \label{sec:2dsgmass}

\begin{figure}
\begin{center}
\includegraphics[width=1.0\columnwidth]{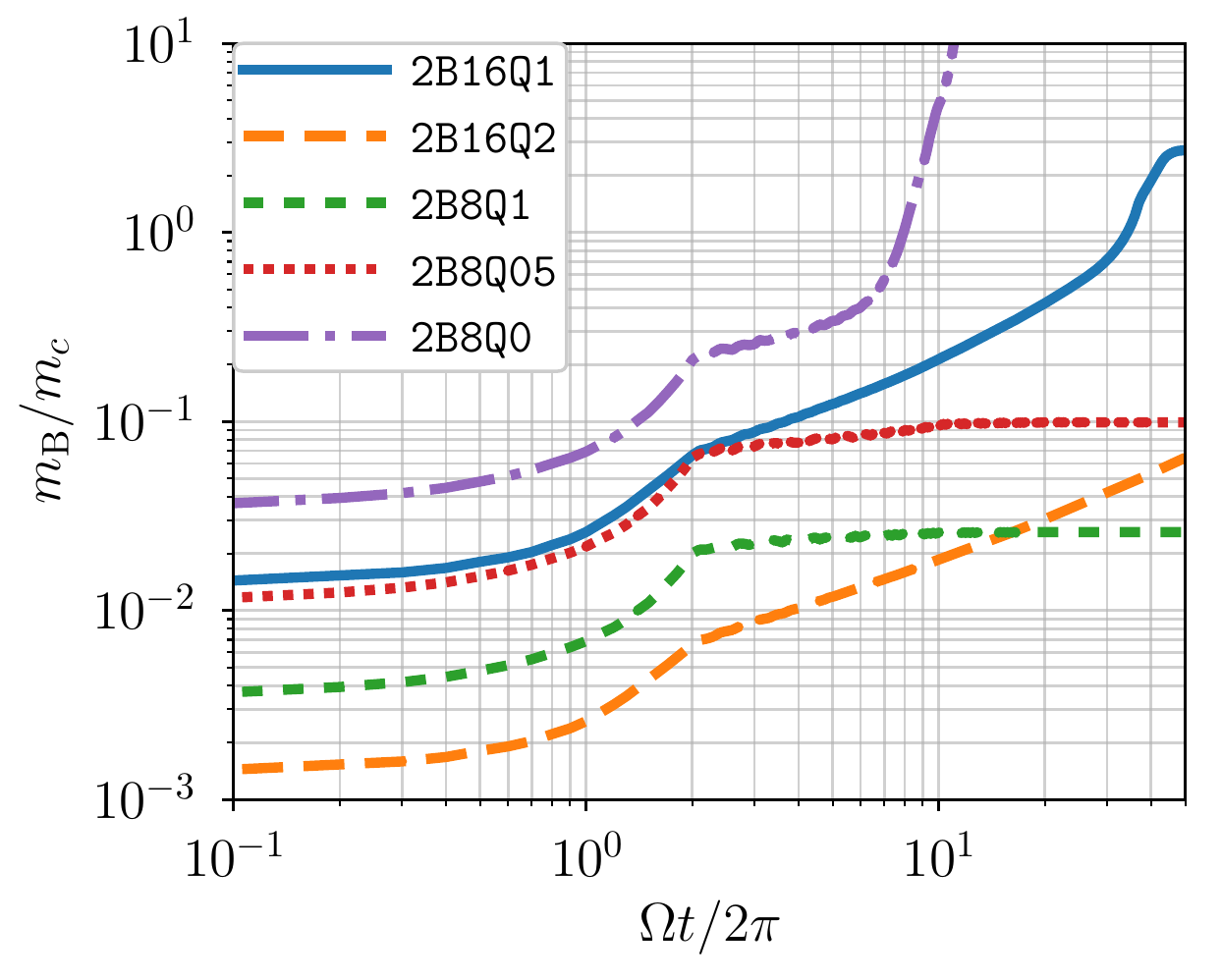}
\caption{Gas mass $m_{\mathrm{B}}$ contained inside the Bondi sphere of the core relative to the core mass $m_c$ as a function of time for the 2D axisymmetric self-gravitating simulations listed in \autoref{tab:2dsg}. The core mass increases up to its nominal value $m_c$ over the first $\Omega t / 2\uppi \leq 2$ orbital times. \label{fig:sgmdot2d}}
\end{center}
\end{figure}

By comparing the background density $\rho_{\infty}$ to the critical value $\rho_{\mathrm{critical}}$, one can predict whether hydrostatic equilibria exist in 1D (rightmost column of \autoref{tab:2dsg}). To test whether this prediction holds in 2D, we draw on Fig. \ref{fig:sgmdot2d} the evolution of the gas mass $m_{\mathrm{B}}$ contained inside the Bondi sphere of the core in each self-gravitating 2D simulation.

The parameters of runs \texttt{2B8Q05} and \texttt{2B8Q1} allow 1D hydrostatic equilibria. In both cases, the Bondi mass $m_{\mathrm{B}}$ converges to a constant value. We can compare it to the envelope mass obtained by integrating the semi-analytic density profiles at $\rho_{\infty} = \rho_{\mathrm{critical}}$ from $r_c$ to $\rB$. For $\rB/r_c=8$ and an outer boundary $\rO/r_c=32$, the critical Bondi mass is $3.32\times 10^{-1} m_c$. Both \texttt{2B8Q05} and \texttt{2B8Q1} indeed converge to a Bondi mass smaller than this critical value. 

With its larger background density, the envelope of run \texttt{2B8Q0} admits no 1D equilibrium. After increasing the mass of the core over the first two orbital times, the envelope mass keeps increasing. After the Bondi mass reaches $m_{\mathrm{B}}/m_c\approx 0.4$ at $\Omega t/2\uppi\approx 6$, the envelope transits to a phase of enhanced mass accretion. The Bondi mass keeps increasing beyond $10$ times the final value in run \texttt{2B8Q1}, and beyond $10^{1/2}$ times the final value in run \texttt{2B8Q05}. The absence of near-linear scaling of $m_{\mathrm{B}}$ with $\rho_{\infty}$ indicates that the envelope of run \texttt{2B8Q0} does not converge toward an equilibrium. In this case only, we stopped the simulation when $m_{\mathrm{B}}/m_c=10$. 

With a larger core mass, the envelopes of runs \texttt{2B16Q1} and \texttt{2B16Q2} are also expected to collapse in 1D. The critical Bondi mass is $2.04\times 10^{-2} m_c$ in this case. Both runs cross this threshold and keep accreting mass until the end of the simulation. In the case of \texttt{2B16Q1}, we note a change of the slope $\partial_t m_{\mathrm{B}}$ at $\Omega t/2\uppi \approx 36$, when the gas mass $m_{\mathrm{B}}$ becomes comparable to the core mass. Whether this transition in accretion rate --- and the one observed in run \texttt{2B8Q0} --- is related to a 1D dynamical collapse is examined in the following section. 

\subsubsection{From core to gas-dominated envelope} \label{sec:2drunaway}

\begin{figure}
\begin{center}
\includegraphics[width=1.0\columnwidth]{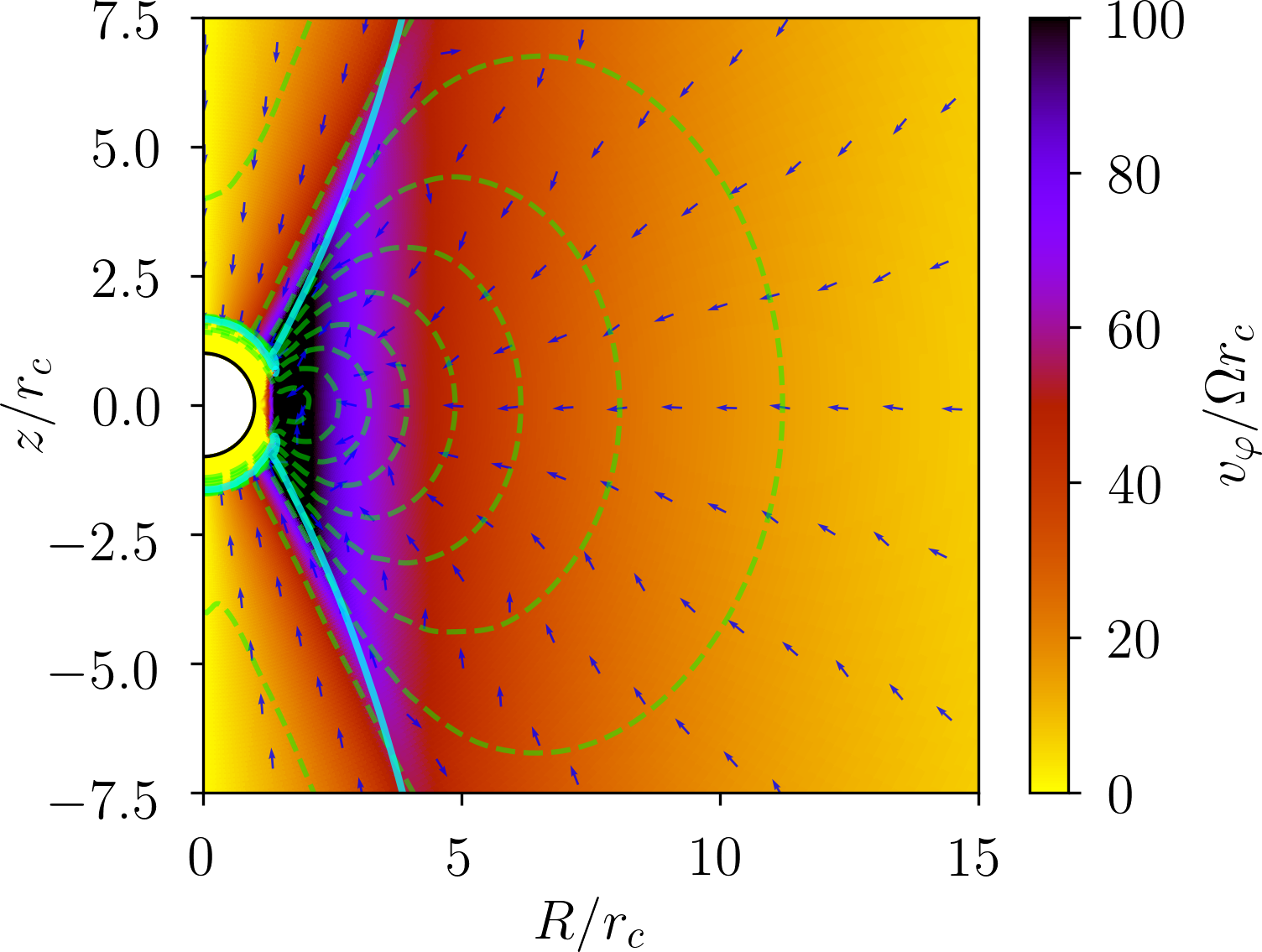}
\caption{Toroidal velocity (\emph{color map}), poloidal mass flux (\emph{blue arrows, orientation only}), mass density (\emph{dashed green contour lines}) and sonic surface for the poloidal velocity (\emph{solid cyan contours}) after time-averaging over $10\Omega^{-1}$ in the simulation \texttt{2B16Q1}. \label{fig:2dvyrhovpsg}}
\end{center}
\end{figure}

Fig. \ref{fig:2dvyrhovpsg} shows the structure of the flow in run \texttt{2B16Q1}, averaged over $\Omega t / 2\uppi \in \left[30,35\right]$. This interval corresponds to the beginning of the enhanced accretion phase on the top right corner of Fig. \ref{fig:sgmdot2d}. We verified that the figure does not change qualitatively when averaging later in the simulation. We also obtained a qualitatively similar picture when averaging the flow over the last two orbital times in run \texttt{2B8Q0}. 

As on Fig. \ref{fig:2dvyrhovp}, the toroidal velocity is larger near the midplane and close to the core. Let $v_{g} \equiv \sqrt{R\partial_R \Phi}$ denote the toroidal velocity required for fully rotational support in the midplane. The ratio $v_{\varphi}/v_{g}=1$ at $1.6\,r_c$, decreasing to $0.5$ at $5.6\,r_c$ and $0.33$ at $9\,r_c$. With rotation dominating the radial momentum balance, the gas density deviates significantly from a spherical, hydrostatic distribution. The density isocontours form lobes anchored in the midplane at $R\simeq r_c$, delimiting a torus of gas orbiting around the core. 

Unlike Fig. \ref{fig:2dvyrhovp}, the poloidal velocity is converging toward the core in the entire plane of Fig. \ref{fig:2dvyrhovpsg}. However, the inflow velocity is supersonic only in the polar accretion cone delimited by the sonic surface. The inflow velocities measured in the midplane remain less than $1$ per cent of the sound speed, so this figure does not depict a global collapse on dynamical timescales. As more mass accumulates on the core, the gravitational potential of the gas becomes deeper. The envelope must therefore contract to maintain radial momentum balance in its own gravitational well. 

\begin{figure}
\begin{center}
\includegraphics[width=1.0\columnwidth]{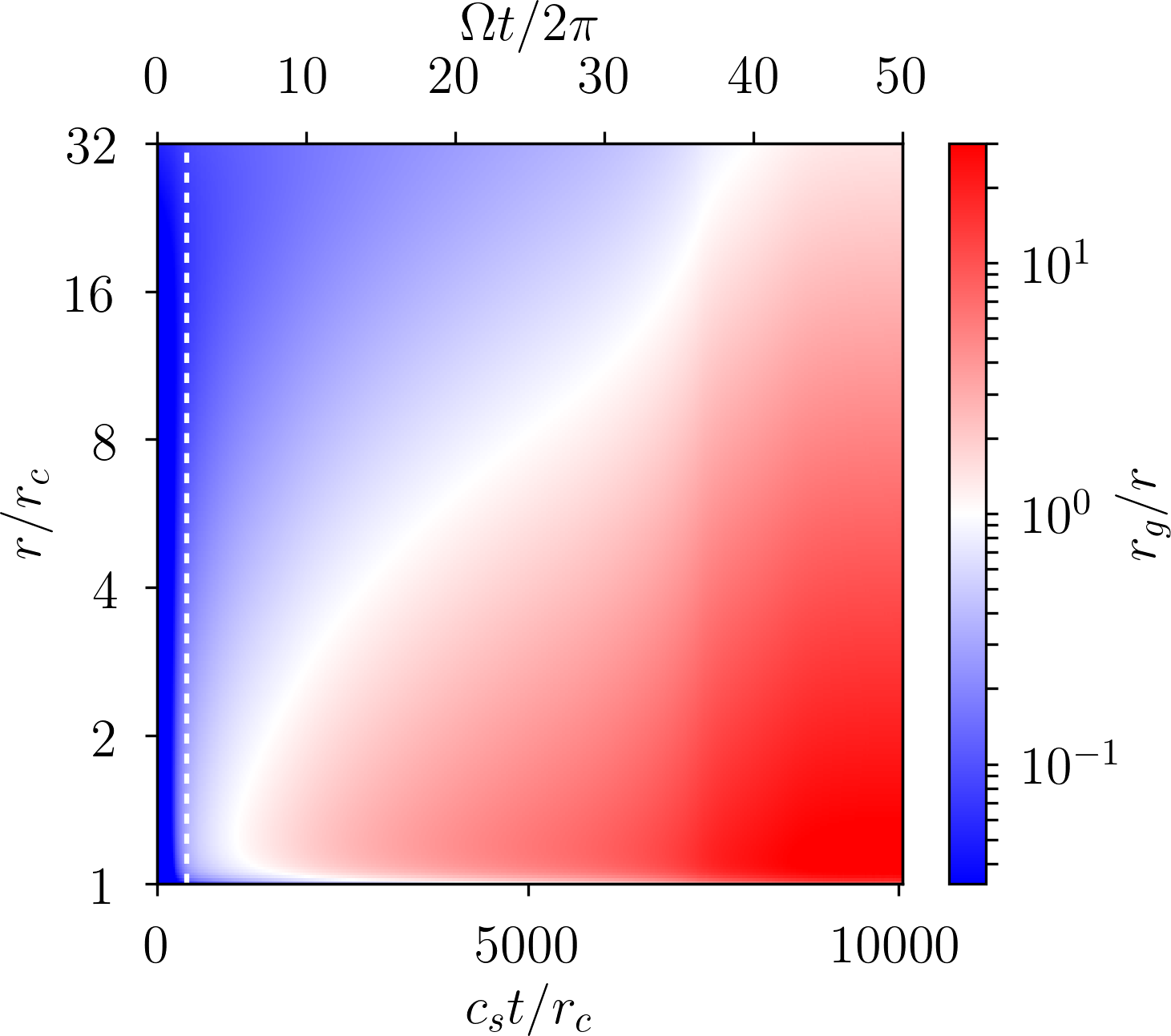}
\caption{Space-time map of the gas gravitational radius $r_g/r$ as defined by \eqref{eqn:rg} relative to the local radius in run \texttt{2B16Q1}. The vertical dashed line marks the time $\Omega t / 2\uppi=2$ when the core mass is fully set. \label{fig:2sgrg}}
\end{center}
\end{figure}

To show the role of the gas gravity more directly, we define the gravitational radius of the cumulative gas mass
\begin{equation}
r_g\left(r\right)\equiv \frac{G \left[m(r)-m_c\right]}{c_s^2}. \label{eqn:rg}
\end{equation}
Fig. \ref{fig:2sgrg} shows the evolution of the gas gravitational radius $r_g/r$ in the simulation run \texttt{2B16Q1}. At $\Omega t / 2\uppi \gtrsim 5$, the gravitational radius of the gas encloses the inner parts of the envelope. The gravitational radius increases in time due to mass accretion, reaching the core's Bondi radius $r_g/r_c=16$ at $\Omega t / 2\uppi \approx 36$. This time also marks the change of accretion rate on Fig. \ref{fig:sgmdot2d}. After $\Omega t / 2\uppi \gtrsim 40$, the gas is bound and attracted toward the core by its self-gravity mainly. 

\subsection{Discussion of 2D models}

Adding rotation leads to the spontaneous formation of a circulatory flow through the envelope. With the parameter space explored in our isothermal simulations, we find preferentially a polar inflow of gas towards the core and an equatorial outflow. This pattern seems robust to boundary effects since it was reported in three-dimensional simulations including the stratification and differential rotation of the background disk \citep[e.g.,][]{tanigawa12,fung15}. 

In the non-gravitating limit, the polar inflows become supersonic when the Bondi radius of the core $\rB/r_c \gtrsim 16$, in agreement with \citet{bethune19b}. The supersonic inflows shock on the inner parts of the envelope, dissipating momentum and allowing mass accretion. Given the limited integration time of our simulations, we can only speculate that these envelopes will converge to a steady state (finite mass) on longer time scales, when the polar shocks expands up to the smooth transsonic point as on Fig. \ref{fig:rhortnosg}. 

When accounting for the gas gravity, we observe a transition to enhanced gas accretion in runs \texttt{2B8Q0} and \texttt{2B16Q1}, as expected from 1D models. In this phase, the envelope mass can increase without bounds, the accretion rate being only restricted by the available gas at the outer radial boundary. However, the accretion rate over the interval $\Omega t / 2\uppi \in \left[2,20\right]$ in run \texttt{2B16Q1} is still unaffected by the gas gravity. This is apparent from the curve of run \texttt{2B16Q2} on Fig. \ref{fig:sgmdot2d}, which has the same slope as run \texttt{2B16Q1} although it is ten times less massive. When run \texttt{2B16Q1} starts accreting at an enhanced rate at $\Omega t / 2\uppi \approx 36$, the envelope mass exceeds the 1D critical mass by a factor $\approx 40$. The transition to a phase of enhanced accretion instead starts when the gas mass becomes comparable to the core mass. 

We do not observe a dynamical collapse of the entire envelope in the 2D simulations of Table \ref{tab:2dsg}. Instead, the accretion flow is limited to the polar regions and a rotationally-supported circumplanetary disk forms near the midplane. We do expect a spherical collapse in the non-rotating limit $\Omega r_c / c_s \rightarrow 0$. Even if a fluid element carries a specific angular momentum $l\simeq R v_{\varphi}$, the acceleration $v_{\varphi}^2/R \simeq l^2/R^3$ alone can balance gravity only up to a limited centrifugal radius. We confirm that a nearly spherical collapse occurs when decreasing $\Omega$ below $1/64$ while maintaining $\Omega r_c / c_s = 1/32$ and $\rB/r_c=16$, i.e., limiting the angular momentum of the gas coming into the computational domain.


\section{Three-dimensional envelopes} \label{sec:3d}

In both the 1D and 2D models, the runaway (enhanced and unbound) accretion phase is controlled by the outer boundary conditions. In reality, the core has a sphere of influence limited by the tidal potential of the star and by the resulting shear flow of the disk. We now examine this situation via 3D self-gravitating simulations of embedded planetary cores. 

\subsection{3D model: rotation and shear}

Let $(x,y,z)$ be cartesian coordinates centered on the planetary core, with $z$ along its rotation axis, $y$ along its orbital trajectory and $x$ along the star-core radius. We consider a small patch of the disk around the core and expand the gravitational potential of the star to second order about the orbital radius of the core \citep{hill78}. Assuming that the circumstellar disk is Keplerian (i.e., neglecting radial pressure gradients), the total potential takes the form
\begin{equation} 
\label{eqn:totpotgrav}
  \Phi = -\underbrace{\frac{3}{2}\Omega^2 x^2}_\text{star}  -\underbrace{\frac{G m_c}{r}}_\text{core} +\underbrace{\Phi_g.}_\text{gas}
\end{equation}
In this patch of the disk, the Keplerian shear flow induced by the star is $v_y = -\left(3/2\right)\Omega x$.

We do not expand the potential of the star in the vertical direction $z$, i.e., we omit the vertical stratification of the disk. We make this choice to facilitate comparisons with the previous 2D results, and to simply subtract\footnote{Otherwise, one would have to solve the Poisson problem $\Delta \Phi_g = 4\uppi G \rho$ for a stratified disk with no planetary core, and then subtract this potential every time the Poisson problem is solved.} a constant value $\rho_{\infty}$ in the modified Poisson problem \eqref{eqn:poisson}. This choice is reasonable when the pressure scale height is large compared to the core radius ($h/r_c \gg 1$). We take $\Omega=1$ and keep $h/r_c = 32$ as in Sect. \ref{sec:2dsg}. 

We extend the computational domain to $\left(r/r_c,\theta,\varphi\right) \in \left[1,128\right]\times\left[0,\uppi\right]\times\left[0,2\uppi\right]$. The radial interval is meshed with $128$ logarithmically spaced grid cells; the $\left(\theta,\varphi\right)$ intervals are meshed with $80 \times 160$ uniformly spaced cells. At the outer radius, we impose the initial conditions of a constant density $\rho_{\infty}$ and a Keplerian shear flow $\left(v_x,v_y,v_z\right) = \left(0,-3\Omega x/2,0\right)$. The inner radial boundary conditions are the same as in the 2D setup, including the homogenized density in the innermost grid shell (see Sect. \ref{sec:2dmodel}). The $\varphi$ boundaries are periodic, and the conditions on the polar axis $\theta=0$ respect the spherical topology of the domain:
\begin{equation*}
\left[\rho,v_r,v_{\theta},v_{\varphi}\right](-\vartheta,\varphi) = \left[+\rho,+v_r,-v_{\theta},-v_{\varphi}\right](\vartheta,\varphi+\uppi)
\end{equation*}
at every radius, and similarly in the opposite hemisphere about $\theta=\uppi\pm\vartheta$. As in Sect. \ref{sec:2d}, the mass of the core is increased up to its nominal value over the first $\Omega t / 2\uppi \leq 2$ orbital times. 

The two main differences with the numerical setup of \cite{bethune19b} are the inclusion of the gas gravity and the omission of the vertical density stratification. The 3D simulations discussed below are listed in \autoref{tab:3dsg}. 

\begin{table}
\begin{center}
\caption{Three-dimensional self-gravitating simulations: label, Bondi radius $r_{\mathrm{B}}/r_c$, Toomre parameter $Q$ as defined by \eqref{eqn:toomre}. \label{tab:3dsg}}
\begin{tabular}{lcc}
Label & $\rB/r_c$ & $Q$ \\
\hline
\verb!3B8Q05! &  $8$ & $10^{0.5}$\\
\verb!3B8Q1!  &  $8$ & $10^{1}$\\
\verb!3B16Q1! & $16$ & $10^{1}$\\
\verb!3B16Q2! & $16$ & $10^{2}$
\end{tabular}
\end{center}
\end{table}

\subsection{Flow structure} \label{sec:3dflow}

\begin{figure}
\begin{center}
\includegraphics[width=1.0\columnwidth]{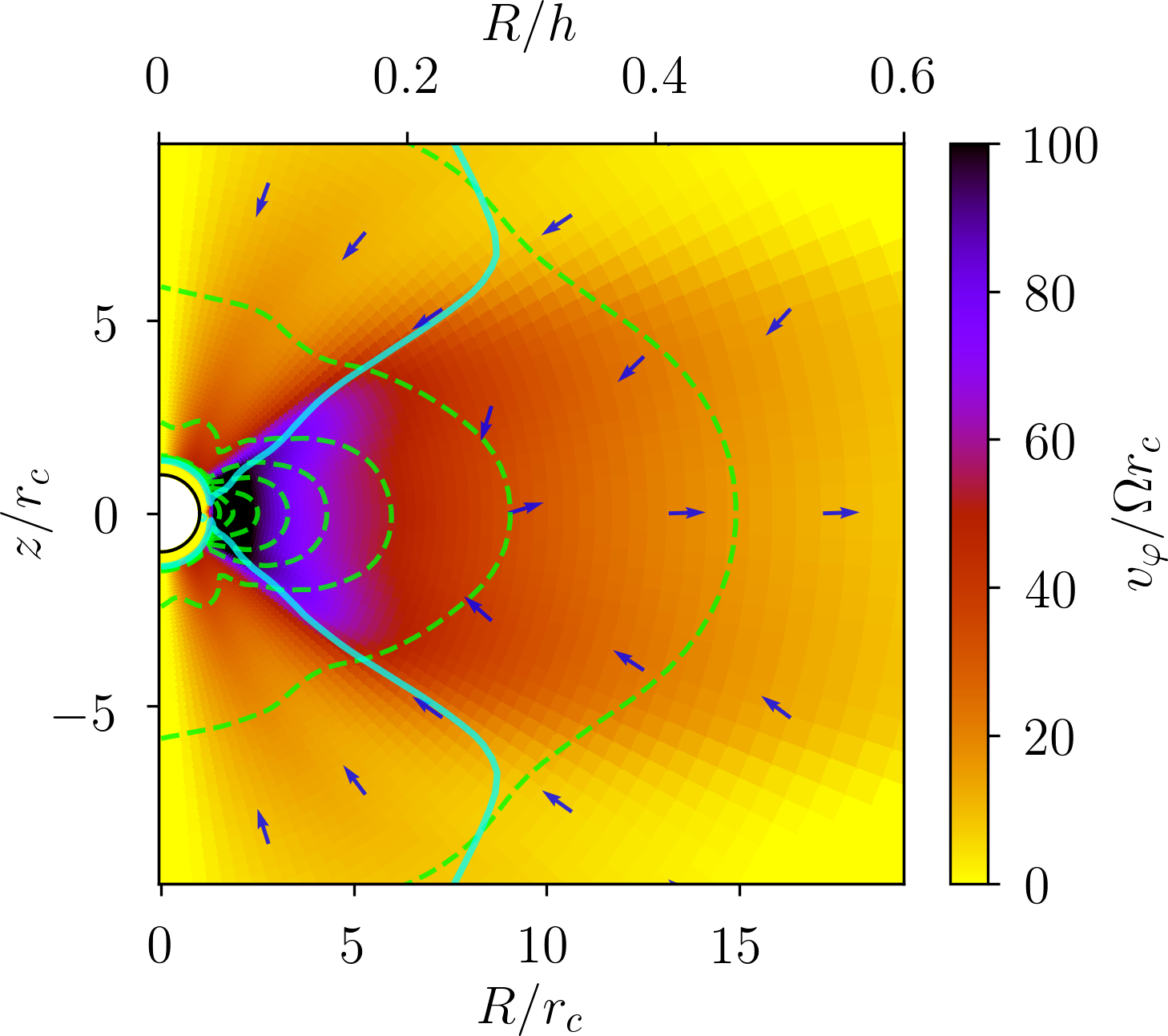}
\caption{Same as Fig. \ref{fig:2dvyrhovpsg} in the equivalent 3D case \texttt{3B16Q1} after azimuthal and time-averaging over $\Omega t / 2\uppi\in\left[5.5,6.5\right]$. \label{fig:3dvyrhovpsg}}
\end{center}
\end{figure}

To compare run \texttt{3B16Q1} with the equivalent 2D case, we average the flow variables in time over one orbital period $\Omega t / 2\uppi\in\left[5.5,6.5\right]$ and in the azimuthal ($\varphi$) direction, and represent them on Fig. \ref{fig:3dvyrhovpsg}. The poloidal mass flux describes a circulatory pattern, with an equatorial outflow and polar inflows shocking close to the core surface. The toroidal velocity reaches $v_{\varphi} / \Omega r_c \gtrsim 100$ in the midplane inside $R \leq 0.2h = 6.4 r_c$. In these inner regions, the density contours form lobes anchored near the surface of the core, delimiting a rotationally-supported circumplanetary disk. One difference with the 2D case of Fig. \ref{fig:2dvyrhovpsg} is the increased opening angle of the accretion cone $\approx 55^{\circ}$, limiting the rotationally-supported envelope to a smaller range of latitudes near the midplane. Another difference with Fig. \ref{fig:2dvyrhovpsg} is the radially limited extent of the circumplanetary disk: the density iso-contours disjoint from the core are restricted to $R\lesssim 6\,r_c$.

\begin{figure}
\begin{center}
\includegraphics[width=1.0\columnwidth]{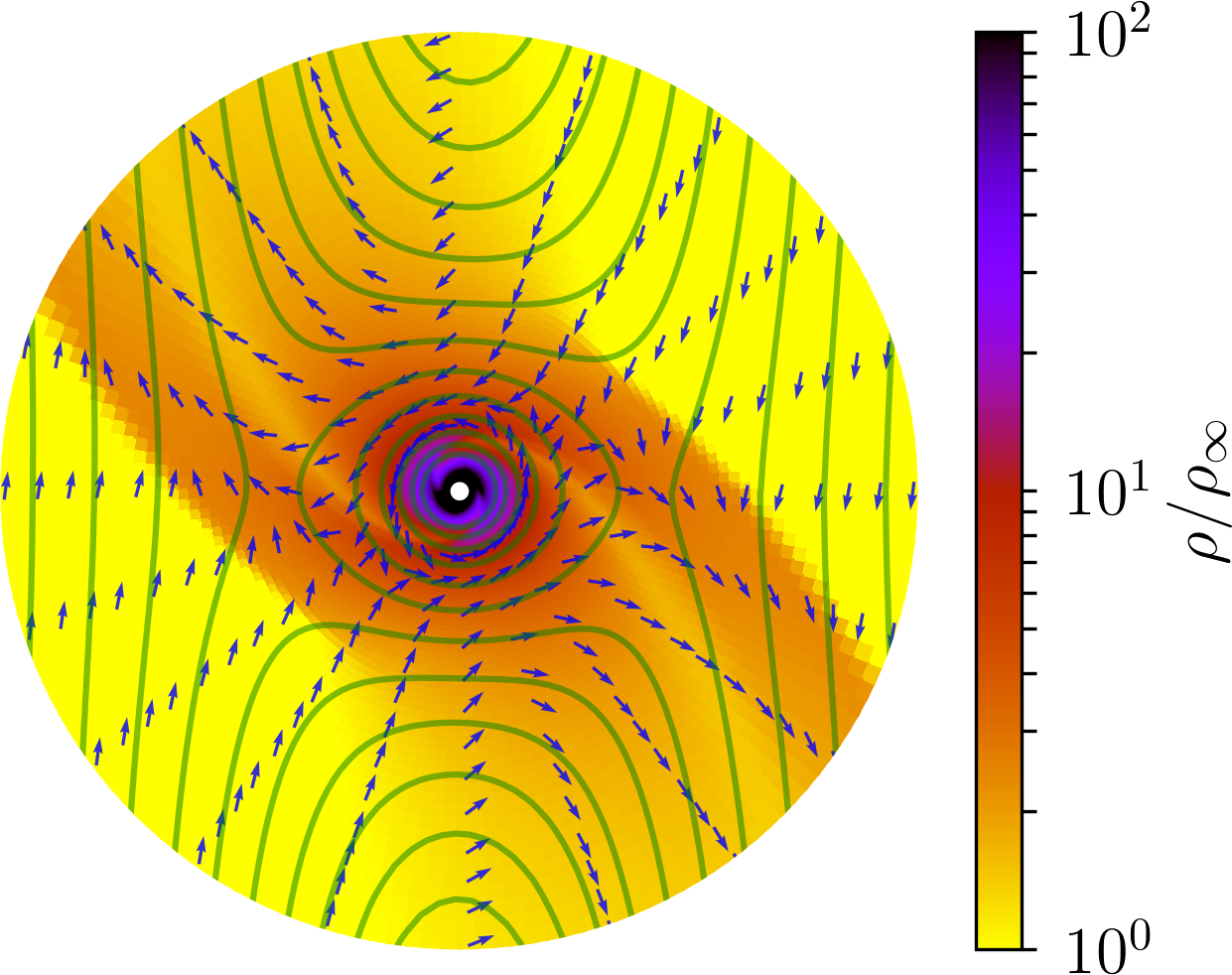}
\caption{Density $\rho/\rho_{\infty}$ (\emph{color map}), velocity (\emph{blue arrows, orientation only}) and total gravitational potential $\Phi$ (\emph{green isocontours}) after time-averaging over $\Omega t / 2\uppi\in\left[5.5,6.5\right]$ in the equatorial plane of run \texttt{3B16Q1}; only the inner $r\leq 2h$ are represented. \label{fig:3dtot11}}
\end{center}
\end{figure}

We show the equatorial flow structure of run \texttt{3B16Q1} on Fig. \ref{fig:3dtot11}. From the isocontours of the gravitational potential, one can identify the different parts of the flow \citep[see for example][]{fung15}. When neglecting the gas gravity, the central region dominated by the potential of the core extends up to the Hill radius $\rH \equiv \left(G m_c / 3 \Omega^2\right)^{1/3}$. With an envelope mass equal to the core mass at this time (see Fig. \ref{fig:sgmdot3d}), the effective Hill radius is only $2^{1/3} \approx 1.26$ larger than in the equivalent non-gravitating case. 

Due to the background shear, the envelope is limited to approximately one pressure scale height in radius. At larger distances $\vert x/h\vert>2/3$, the Keplerian shear flow is supersonic with respect to the core; the density perturbations induced by the core are then transported by spiral density waves into the disk. Despite their large spatial extent, we find no significant contribution of the spiral waves to the gravitational potential of the gas, which remains spherically symmetric in good approximation. 

\subsection{Mass of 3D gravitating envelopes}

\begin{figure}
\begin{center}
\includegraphics[width=1.0\columnwidth]{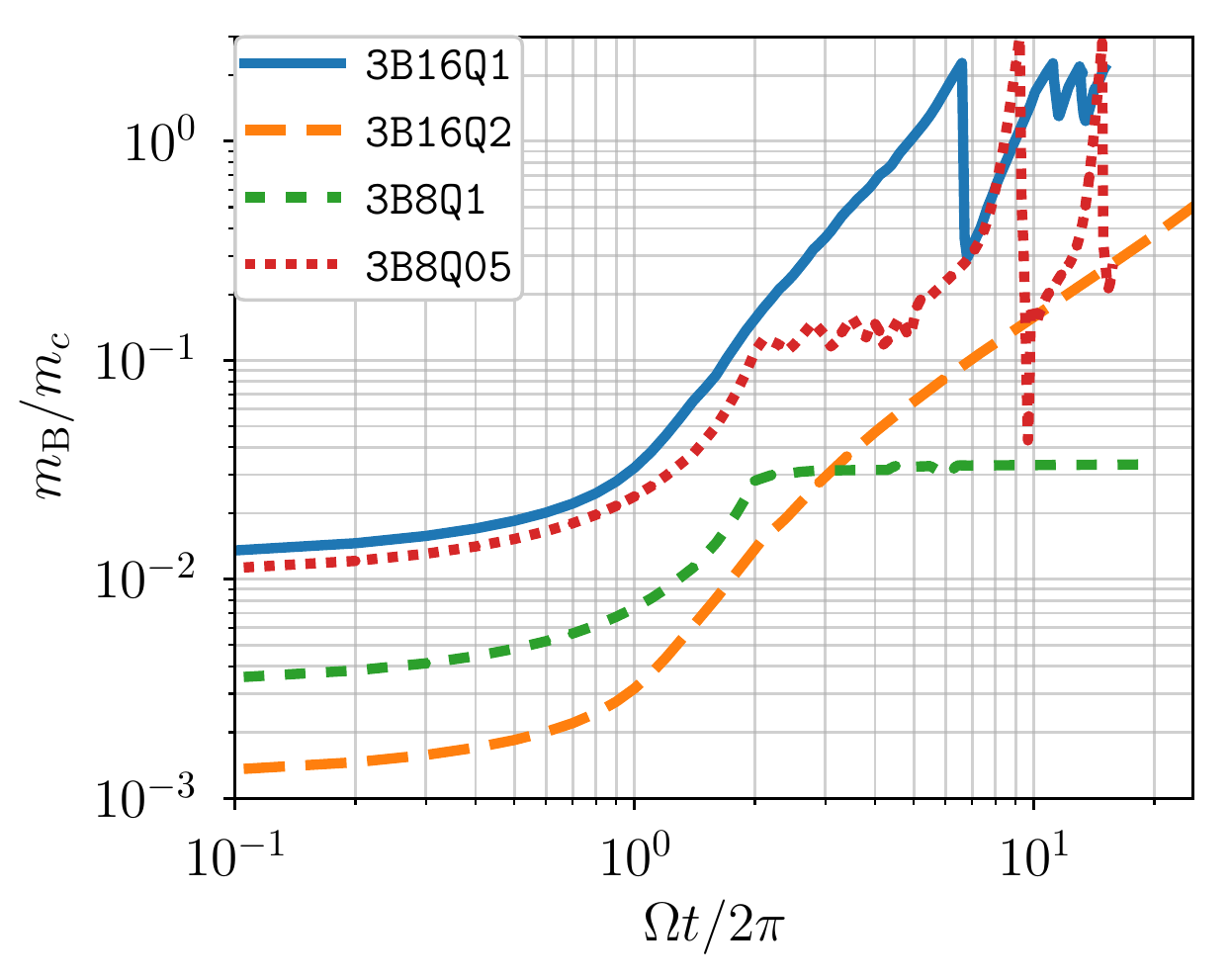}
\caption{Gas mass contained inside the Bondi sphere of the core relative to the core mass $m_c$ as a function of time for the 3D self-gravitating simulations listed in \autoref{tab:3dsg}. The core mass increases up to its nominal value $m_c$ over the first $\Omega t / 2\uppi \leq 2$ orbital times. \label{fig:sgmdot3d}}
\end{center}
\end{figure}

For each 3D simulation listed in \autoref{tab:3dsg}, we integrate the gas mass $m_{\mathrm{B}}$ contained inside the Bondi radius of the core, normalize it by the final mass $m_c$ of the core, and draw its evolution on Fig. \ref{fig:sgmdot3d}. 

Only run \texttt{3B8Q1} has an envelope mass converging to a finite value. The residual accretion rate is approximately $5\times 10^{-6}\,\Omega m_c$, negligible over the duration of the simulation. With a laminar and subsonic poloidal flow (no polar shocks), mass accretion is mainly driven by numerical dissipation in this simulation.

The envelope mass in run \texttt{3B16Q2} increases up to $50$ per cent of the core mass over the integration time of the simulation. Mass accretion is caused by polar shocks as described in Sect. \ref{sec:3dflow}. The mass accretion rate increases from $1.9\times 10^{-2} m_c$ per orbit to $2.6\times 10^{-2} m_c$ per orbit over the interval $\Omega t/2\uppi\in\left[8,24\right]$. 

In the last two simulations \texttt{3B16Q1} and \texttt{3B8Q05}, the combination of a large core mass and/or a large background density leads to the most massive envelopes. Both envelopes accrete gas and become as massive as the core during the simulation. However, these envelopes saturate at $m_{\mathrm{B}}/m_c \lesssim 3$. At $\Omega t/ 2\uppi \approx 6.6$ in run \texttt{3B16Q1} and $9.5$ in run \texttt{3B8Q05}, the envelope mass drops by an order of magnitude over a fraction of orbital time. 

\begin{figure}
\begin{center}
\includegraphics[width=1.0\columnwidth]{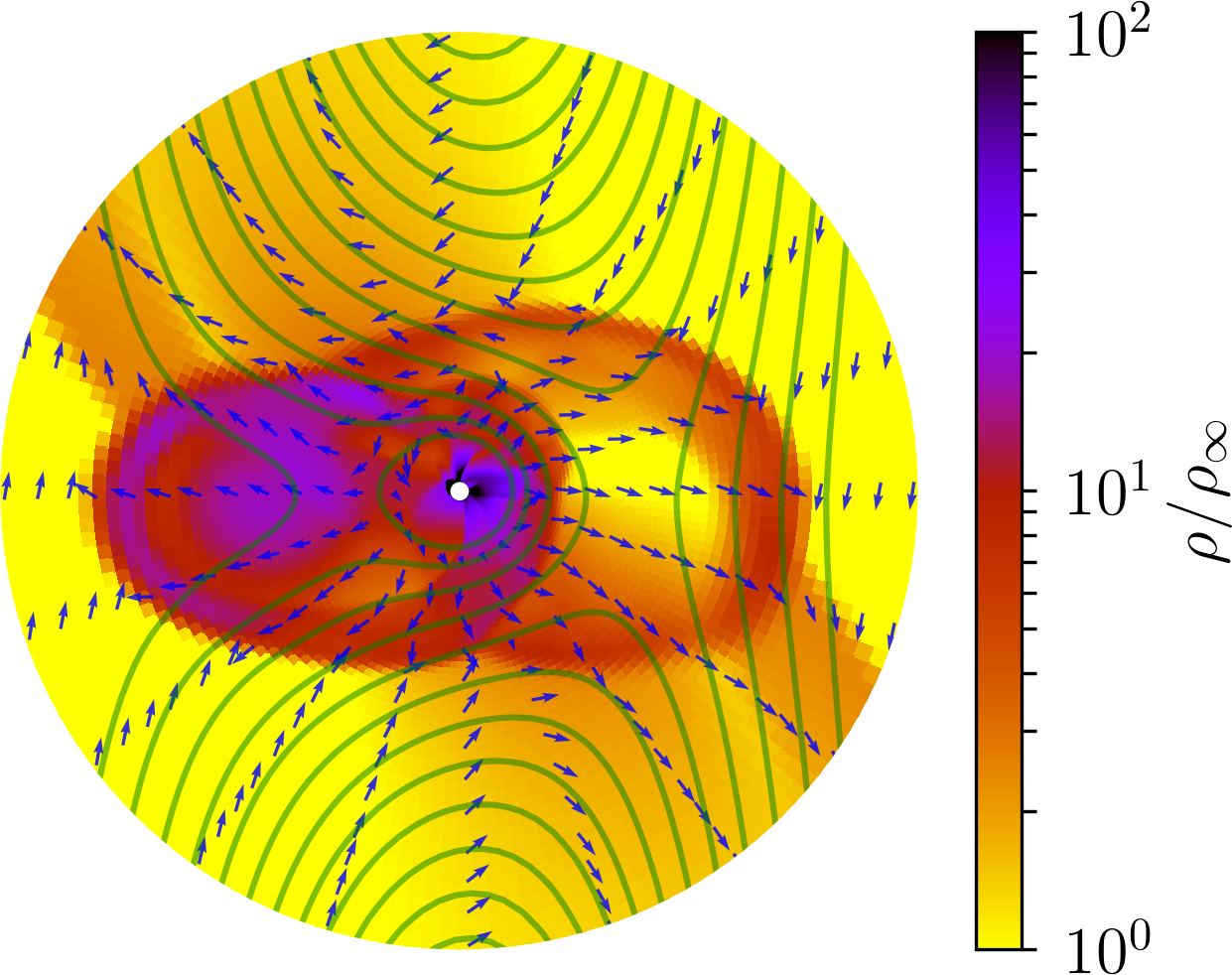}
\caption{Same as Fig. \ref{fig:3dtot11} but at the time $\Omega t/2\uppi = 6.7$ in run \texttt{3B16Q1}, corresponding to the envelope mass drop on Fig. \ref{fig:sgmdot3d}. \label{fig:3dtot14}}
\end{center}
\end{figure}

To understand what caused the envelope mass to drop, we show on Fig. \ref{fig:3dtot14} a snapshot of the equatorial flow in run \texttt{3B16Q1} at $\Omega t / 2\uppi = 6.7$. The density distribution features two arc-shaped shocks enclosing the core. Inside the bubble delimited by these shocks, the gas moves away from the core at supersonic velocity. The gravitational potential is distorted compared to Fig. \ref{fig:3dtot11}. The isocontours of potential are pinched toward the core on the left side of Fig. \ref{fig:3dtot14}, indicating that the blob of gas ejected to the left carries a mass comparable to that of the core.

The envelope of run \texttt{3B8Q05} `explodes' in a similar fashion at $\Omega t / 2\uppi \approx 9.5$. The explosions are preceded by the fragmentation of the gas in the innermost parts of the envelope. The resulting density clumps exert a torque on the surrounding material, throwing most of the envelope mass outside the Hill sphere of the core. To assert whether this fragmentation is caused by Jeans instability, we compute the Jeans length $\lJ^2 \equiv \uppi c_s^2 / G \rho'$. In run \texttt{3B16Q1} at $\Omega t / 2\uppi = 6.5$, the shell-averaged Jeans length goes as low as $0.1\,r_c$ near the core surface, resolved by only $3$ to $5$ grid cells. This length is smaller than the extent of the shocked shell $\approx 0.3\,r_c$ on Fig. \ref{fig:3dvyrhovpsg}, so the envelope could be unstable to radial density perturbations. 

We did not observe gravitational fragmentation in the equivalent 2D simulation \texttt{2B16Q1}. We successfully reproduced the 2D runs \texttt{2B16Q1} and \texttt{2B8Q05} at a reduced resolution of $96\times 96$ cells over $\left( r,\theta\right)$, comparable to the 3D resolution inside $r/r_c\leq 32$. The absence of fragmentation in these low-resolution 2D runs supports that fragmentation is not caused by the lower numerical resolution of our 3D simulations. It also points towards non-axisymmetric disturbances to trigger the envelope fragmentation. Higher resolution 3D simulations will be required to examine this issue when computational resources allow it.

After the explosion, the remaining envelope is less massive than the core again. The envelope remnant settles as a new hydrostatic shell, and the process of polar mass accretion resumes until the next explosion. This leads to the saturated behavior of the envelope mass with time shown on Fig. \ref{fig:sgmdot3d} for runs \texttt{3B16Q1} and \texttt{3B8Q05}.

\section{Discussion} \label{sec:discussion}

\subsection{Comparison with previous works}

\subsubsection{Non-gravitating envelopes}

The 3D simulations are most easily compared with those of \citet{bethune19b}, who used a nearly identical setup. Comparing run \texttt{3B16Q1} with the equivalent non-gravitating run \texttt{H32B16} (see their Figure 10 (e)), the accretion cones are wider, the inflow becomes supersonic higher above the core, and the density contours are more pinched around the midplane. \texttt{3B16Q1} shares more similarities with run \texttt{H32B32} on Figure 10 (f) of \cite{bethune19b}, for which the Bondi radius of the core is twice larger ($\rB/r_c=32$). This is consistent with the fact that the envelope mass equals the core mass at this time in run \texttt{3B16Q1}, see Fig. \ref{fig:sgmdot3d}. In other words, the envelope structure in run \texttt{3B16Q1} is the same as if the gas mass was simply added to the core mass. To allow a more quantitative comparison, we reproduce Fig. \ref{fig:3dvyrhovpsg} for the non-gravitating simulation \texttt{H32B32} of \cite{bethune19b} in Appendix \ref{app:3dnosg}. 

We estimate the size of the circumplanetary disk formed in run \texttt{3B16Q1} as the radius $R_d$ at which the specific angular momentum $R v_{\varphi}$ is maximal. The disk size $R_d \approx 0.32\times \left(1.26 \,\rH\right)$ is remarquably close to the $\rH/3$ predicted by \citet{quillen98} after correcting the Hill radius $\rH$ by a factor $2^{1/3}$ for the enclosed gas mass. This ratio is larger than the $\rH/10$ found by \citet{wang14} in isothermal simulations around intermediate-mass cores, but it is supported by the radiative simulations of \citet{danhenkley2003a} and \citet{ayliffe09b} around high-mass cores. 

Because the accretion flow is restricted to the polar cones, the mass accretion rate should be sensitive to the vertical stratification of the disk. Since we omit the disk stratification in the present study, we expect larger accretion rates compared to those of \citet{machida10,bethune19b}. The accretion rate of $2\times10^{-2} m_c$ per orbit in run \texttt{3B16Q2} is indeed $10^3$ times larger than predicted from equation (13) of \citet{bethune19b}. 

\subsubsection{Self-gravitating envelopes}

\citet{wuchterl3} presented radiative simulations of 1D envelopes at the critical core mass. He found that the core loses most of its envelope by launching an \emph{outflow}. In our 1D isothermal simulations, the `instability' always develops as an \emph{inflow} toward the core. As discussed in Sect. \ref{sec:1ddiscussion} above and in section 3.2 of \citet{wuchterl3}, no hydrostatic equilibria connecting the core to the disk can be found beyond the critical mass. Whether the subsequent evolution is a collapse or an expansion of the envelope should therefore be determined by the outer boundary conditions. In the simulations of \citet{wuchterl3}, the critical solution is taken as initial condition, with no initial velocity. In our case, the gas already flows inward when the core reaches the critical mass. This difference could select the inflow as a favored outcome of our setup. 

Beyond the critical mass, one can still find hydrostatic equilibria for an isolated planetary envelope. \citet{perricameron74}, \citet{mizuno78} and \citet{wuchterl1} found that these equilibria are linearly unstable due to the adiabatic index $\gamma<4/3$ in the H$_2$ dissociation region. We confirm the linear instability of these equilibria in Appendix \ref{app:linstab}, where we also find unstable modes for $\gamma>4/3$. We argue that this linear analysis might not be relevant for embedded planets because of fast reaction (e.g., free-fall of the surrounding disk) happening at the critical mass. 

\cite{ayliffe12} described the hydrodynamics of gravitating planetary envelopes in 3D radiative simulations. The envelope of their most massive core ($33$M$_{\oplus}$ in model J) undergoes a hydrodynamic collapse, after which it settles into a new equilibrium and resumes gas accretion. We do not observe a collapse of the hydrostatic inner envelope in our multi-dimensional simulations, even when they enter an enhanced accretion phase. The sudden contraction of the inner envelope in model J of \cite{ayliffe12} might be caused by an opacity drop as the temperature rises near the core, leading to a more isothermal (steeper) density profile. 

\subsection{From 1D to 3D}

In 1D models the radial extent of the envelope must be prescribed a priori. We showed in Sect. \ref{sec:critmass} that the location of this boundary weakly affects the value of the critical mass when $\rB/r_c\geq 8$. In 3D the envelope is limited by the background shear to roughly one pressure scale $h\equiv c_s/\Omega$ around the core. The 2D axisymmetric simulations extending to $h$ should therefore be comparable to the 3D simulations regarding the core-nucleated instability. 

From 1D to 2D axisymmetric, we still find a transition to runaway (enhanced and unbound) gas accretion. However, mass accretion is initially driven by transsonic inflows with no role of the gas gravity. The mass accretion rate increases only when the envelope mass becomes comparable to the core mass, which can be orders of magnitude larger than the critical mass predicted in 1D static models. If the mass accretion rate is proportional to the mass of the planet as in non-gravitating simulations \citep{machida10,bethune19b}, then the planet mass should increase exponentially in time --- as long as the disk can provide this material. 

From 2D axisymmetric to 3D, we can compare the evolution of the envelope mass on Fig. \ref{fig:sgmdot2d} and Fig. \ref{fig:sgmdot3d} respectively. The mass accretion rate in run \texttt{3B16Q2} is twenty times larger than in the equivalent 2D run \texttt{2B16Q2}. This is related to the different properties of the background flow. In 2D, the incoming gas has a prescribed angular momentum as if the outer boundary was in solid body rotation. In 3D, the shear flow has a different vorticity distribution \citep{krumholz2005}, resulting in wider accretion cones and larger accretion rates. Unlike their 2D analogues, the 3D accretion rates of runs \texttt{3B16Q1} and \texttt{3B16Q2} do not significantly increase when the ratio $m_{\mathrm{B}}/m_c \gtrsim 50$ per cent. The fragmentation of the inner envelope happens before a phase of accelerated growth can be clearly identified. 

\subsection{Gas thermodynamics}

The envelopes presented by \cite{ayliffe12} all accrete gas at an accelerating rate when the envelope mass becomes comparable to the core mass (see their Figure 1). The dust opacity --- which might vanish in the vicinity of the core \citep{podolak03,movshovitz10} --- controls the efficiency of radiative cooling, and therefore the envelope contraction and accretion \citep[see][Figure 2]{ayliffe12}. If radiative cooling only affects the timescale of mass accretion, e.g., neglecting envelope recycling \citep{ormel2,kurokawa18}, then our simulations could be appropriate in late stages of gas accretion onto high-mass cores \citep[$\geq 15\,$M$_{\oplus}$ in model A of][]{ayliffe12}.  

The envelope mainly accretes gas through polar inflows, with a negligible contribution from the circumplanetary disk to the mass accretion rate \citep{machida10}. For an isothermal gas, the inflows lose momentum by shocking on the inner envelope. For an adiabatic gas, the momentum would first be converted into heat, which would then have to be radiated away \citep{szulagyi14}. Calculations of the shock radiation efficiency point toward isothermal conditions \citep{marleau17,marleau19}. The mass accretion rate parametrized by \citet{machida10} and \citet{bethune19b} could therefore be used to determine the time required for gas gravity to become important after a supersonic inflow develops. 

Regarding the fragmentation of the envelope in runs \texttt{3B16Q1} and \texttt{3B8Q05} (see Fig. \ref{fig:3dtot14}), it is made possible by the somewhat unrealistic isothermal equation of state. Isothermal envelopes have the largest mass and the steepest density profile given a background temperature, making them most easily prone to Jeans instability near the core surface. In case of instability, the gravitational collapse would be limited for an adiabatic contraction of the envelope, and eventually regulated by the efficiency of radiative cooling.

\section{Summary and perspectives}

We studied the hydrodynamics of embedded planetary envelopes in the regime where the gravity of the gaseous envelope becomes comparable to the gravity of the solid core. We focused on isothermal envelopes and considered three models of increasing complexity:
\begin{enumerate}
\item a 1D model for spherically symmetric envelopes, helping us investigate the nature of the core-nucleated instability through hydrostatic and hydrodynamic calculations;
\item a 2D model for axisymmetric envelopes, allowing a rotationally-supported circumplanetary disks to form by conservation of angular momentum;
\item a 3D model including the tidal potential of the star, where the planetary core is embedded in the differentially rotating circumstellar disk, but omitting the vertical stratification of the disk. 
\end{enumerate}

We summarize our main conclusions as follows:
\begin{itemize}
\item In spherically-symmetric envelopes, the core-nucleated instability corresponds to the absence of equilibrium connecting the core to the ambient conditions in the circumstellar disk. The following reaction can be a contraction or an expansion of the envelope until momentum balance is satisfied again. 
\item Including rotation, the formation of a circumplanetary disk does not prevent the transition to runaway \edt{(accelerated and unbound)} gas accretion; it only restricts the accretion flow to the polar cones, where the gas has a negligible angular momentum with respect to the core. 
\item In rotationally-supported envelopes, the accelerated accretion phase starts when the envelope mass becomes comparable to the core mass, irrespective of the critical mass computed in 1D models. 
\item Because most of an isothermal envelope mass is accumulated at the surface of the core, the flow structure is the same as if the gas mass was simply added to the core mass in a non-gravitating medium. The mass of 3D isothermal envelopes saturate around a few core masses due to fragmentation, \edt{preventing their unlimited growth}. 
\end{itemize}

The main shortcoming of this study is the extremely simplified treatment of the gas thermodynamics. Future works should address this issue by progressively accounting for radiative energy transport and gas chemistry, expanding the results of \citet{ayliffe12}. To gain some predictive power, this study should also be repeated by embedding the core within a global disk model. This step is necessary to understand the outcome of runaway gas accretion after the planet opens a gap in the disk \citep{ginzburgchiang19} and the disk eventually disperses \citep{alexander13}.

\section*{Acknowledgements}

Financial support of this work by the Isaac Newton Trust and the Department of Applied Mathematics and Theoretical Physics is gratefully acknowledged. I thank Richard Booth for his early suggestions, and the anonymous referee for his constructive comments that improved the quality of this paper. The results presented here were obtained before the internship of Sabina Sagynbayeva, whom I co-supervised with Roman R. Rafikov on 1D simulations of self-gravitating isothermal envelopes. I thank both of them for helping me develop more lucidity on this topic.


\bibliographystyle{mnras}
\bibliography{biblio} 



\appendix

\section{Numerical accuracy of the Poisson solver} \label{app:errors}

To solve the Poisson problem \eqref{eqn:poisson} in \textsc{pluto}, we use the parallel solver described in Appendix B of \cite{bethune19a}. The second-order finite difference discretization of the Laplacian operator might introduce numerical errors when the density distribution becomes steep. To verify that the Poisson solver performs well in the simulations presented above, we compute the error
\begin{equation} \label{eqn:errpoisson}
\epsilon \equiv \frac{\Delta \Phi_g}{4\uppi G \rho'} - 1
\end{equation}
in the 1D self-gravitating simulation of Fig. \ref{fig:rhortwithsg}. With a density contrast of $10^8$, this simulation is our most defavorable case regarding numerical accuracy. 

\begin{figure}
\begin{center}
\includegraphics[width=1.0\columnwidth]{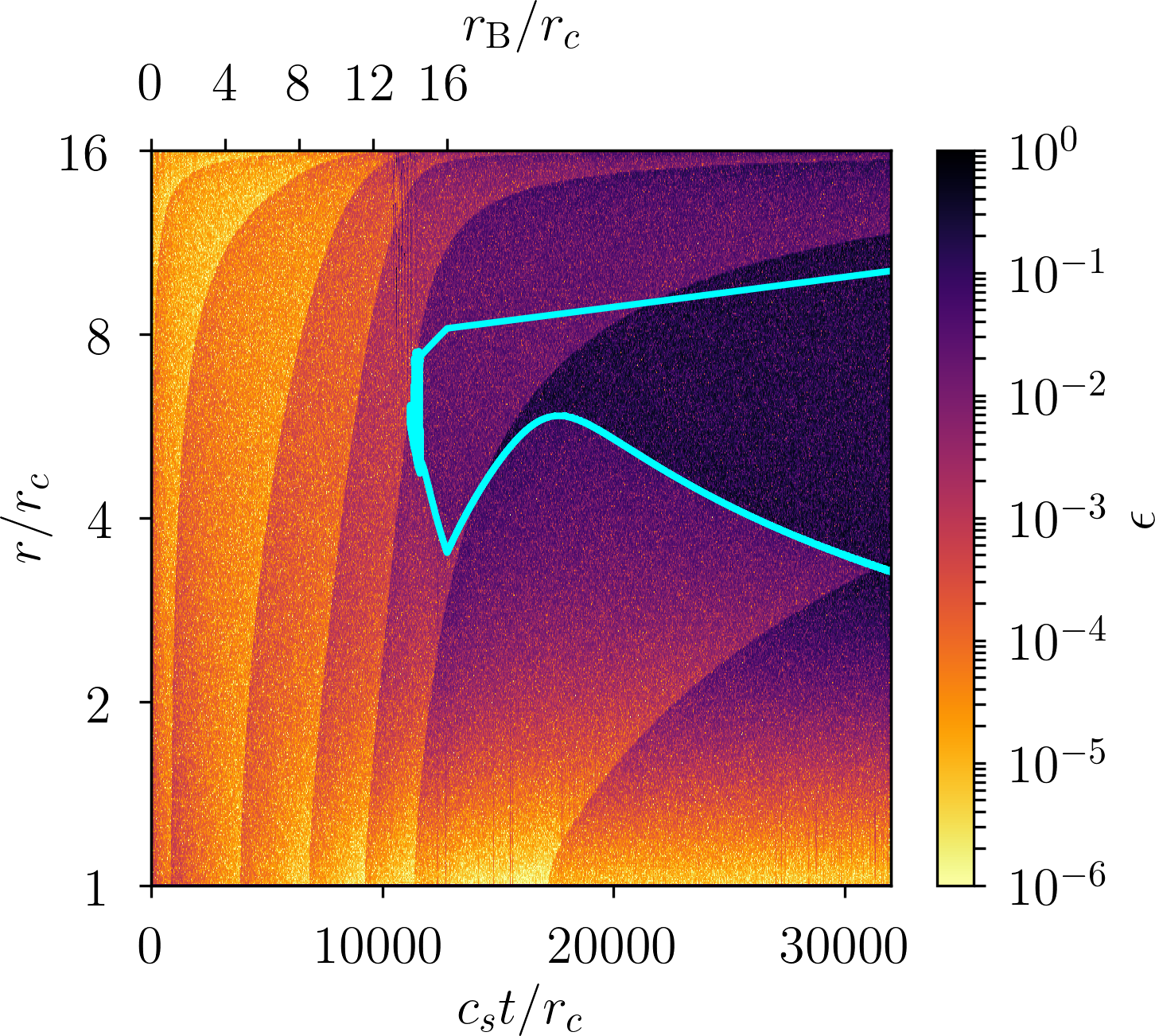}
\caption{Relative error of the Poisson solver as defined by \eqref{eqn:errpoisson} in the 1D self-gravitating simulation of Fig. \ref{fig:rhortwithsg}. The cyan contour line marks the sonic points $v_r=-c_s$. \label{fig:errpoisson}}
\end{center}
\end{figure}

The error $\epsilon$ in this simulation is shown on Fig. \ref{fig:errpoisson} as a function of time and radius. The maximal error is less than $10^{-2}$ before the inflow becomes supersonic, and later between $0.1$ and $0.3$ upstream of the shock. The absolute error on $\Delta\Phi_g$ is continuous across the shock, so the apparent jump in $\epsilon$ only comes from the jump in $\rho'$. After a shock forms, the Poisson problem is dominated by the innermost regions, where the Laplacian $\Delta\Phi_g$ remains overall resolved to per cent accuracy. 

\section{Three-dimensional non-gravitating simulation} \label{app:3dnosg}

\begin{figure}
\begin{center}
\includegraphics[width=1.0\columnwidth]{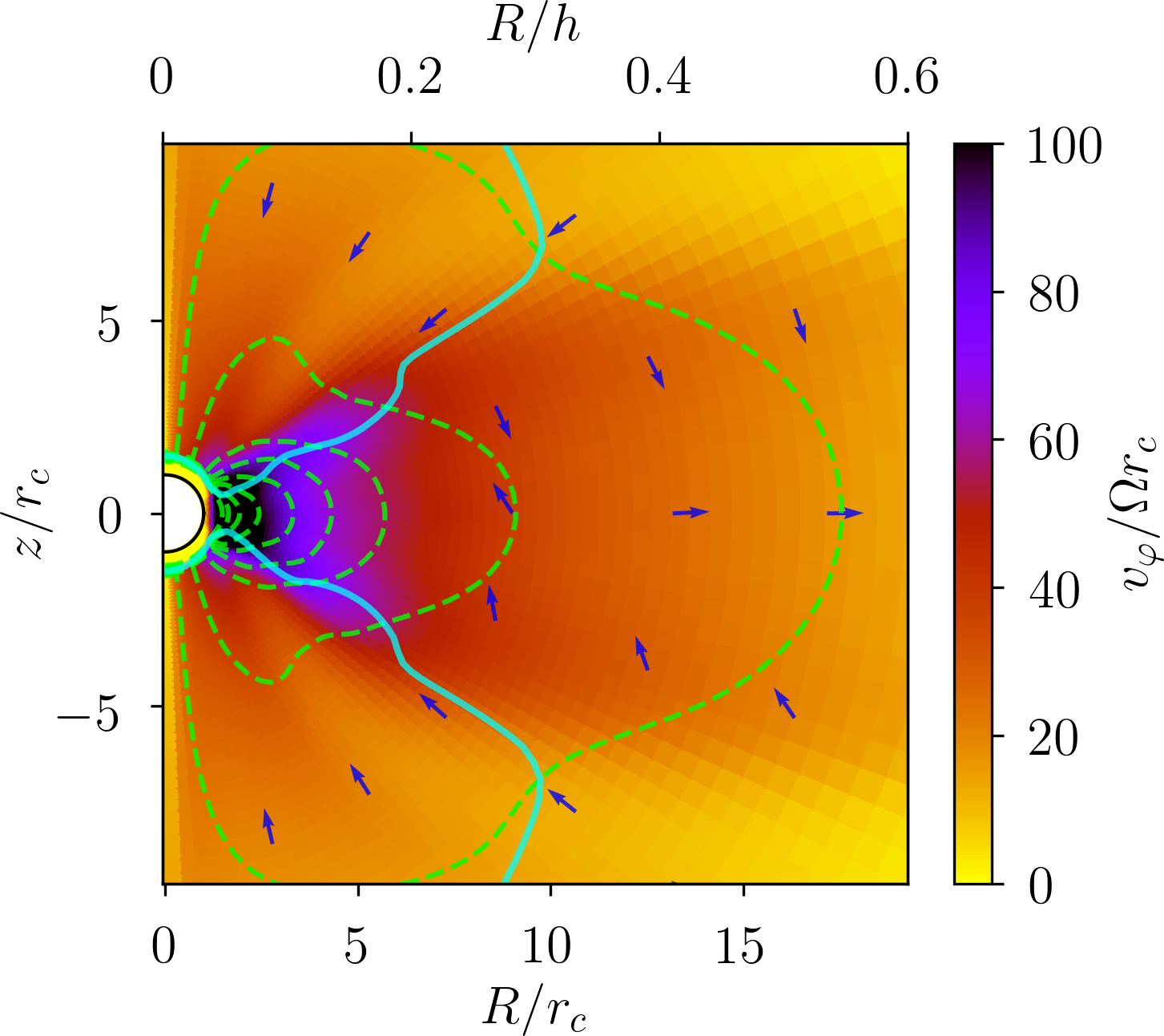}
\caption{Non-gravitating 3D isothermal simulation \texttt{H32B32} of B\'ethune \& Rafikov (2019b) with a Bondi radius $\rB/r_c=32$ and including vertical stratification on a pressure scale $h/r_c = 32$, to be compared with Fig. \ref{fig:3dvyrhovpsg}. \label{fig:i3dp32b32}}
\end{center}
\end{figure}

Fig. \ref{fig:i3dp32b32} shows the time and azimuthally averaged flow structure in run \texttt{H32B32} of \citet{bethune19b}. Unlike run \texttt{3B16Q1} shown on Fig. \ref{fig:3dvyrhovpsg}, this envelope is non-gravitating, the background disk is vertically stratified on a pressure scale $h = 32\,r_c$, and the core is twice more massive with $\rB=32\,r_c$. 

In comparison with Fig. \ref{fig:3dvyrhovpsg}, the density iso-contours are pinched toward the core near the polar axis. This feature is caused by the low-density polar inflows; it is the only region where the vertical stratification of the disk affects the envelope. Near the midplane, the density iso-contours form lobes with a similar aspect ratio, also delimiting a circumplanetary disk. The toroidal velocity reaches similar amplitudes $v_{\varphi} \lesssim 100\,\Omega r_c$ in the midplane.

\section{Linear stability of self-gravitating envelopes} \label{app:linstab}

We examine the linear stability of the 1D self-gravitating equilibria obtained in Sect. \ref{sec:semianal}. These equilibria represent isolated / unbound planetary envelopes. They cannot satisfy arbitrary condition on $\rho$ away from the core, so they cannot be continuously connected to the ambient disk in general. 

We set the outer radius $\rO/r_c=16$, the Bondi radius $\rB/r_c=8$, and the background temperature via $P(\rO)=\rho(\rO) c_s^2$. Given a polytropic exponent $\gamma$, we find the background density beyond which the boundary condition $\rho(\rO)=\rho_{\infty}$ cannot be satisfied anymore. Beyond this point, we can still parametrize the solutions by their envelope mass $m_g\equiv \int_{r_c}^{\rO} 4\uppi r^2 \rho \,\dd r$ relative to the mass $m_{\mathrm{critical}}$ integrated at the threshold $\rho_{\infty}=\rho_{\mathrm{critical}}$. With this parametrization, the residual error on the solutions is $10^{-12}$ even above the critical mass. 

Given an equilibrium profile $\bar{\rho}$, the linearized equations for the perturbed density, velocity and gravitational acceleration $\left(\rho',v_r',g_r'\right)$ for a polytropic gas with $P=\kappa\rho^{\gamma}$ are:
\begin{align}
  \partial_t \rho' &= - \frac{1}{r^2}\frac{\partial}{\partial r}\left[r^2 \bar{\rho} v_r'\right],\label{eqn:dtrho}\\
  \partial_t v_r'  &= - \frac{1}{\bar{\rho}} \frac{\partial}{\partial r} \left[\gamma\kappa \bar{\rho}^{\gamma-1} \rho'\right] + \frac{\rho'}{\bar{\rho}^2} \frac{\partial}{\partial r} \left[\kappa \bar{\rho}^{\gamma}\right] + g_r', \label{eqn:dtvr}\\
  \partial_t g_r'  &= 4\uppi \bar{\rho} v_r'. \label{eqn:dtgt}
\end{align}
We look for eigenmodes satisfying $\rho'(\rO)=v_r'(r_c)=g_r'(r_c)=0$, and with a time dependence $\sim \exp\left[\omega t\right]$. We focus on real and positive eigenvalues $\omega$, i.e., unstable modes. 

\begin{figure}
\begin{center}
\includegraphics[width=1.0\columnwidth]{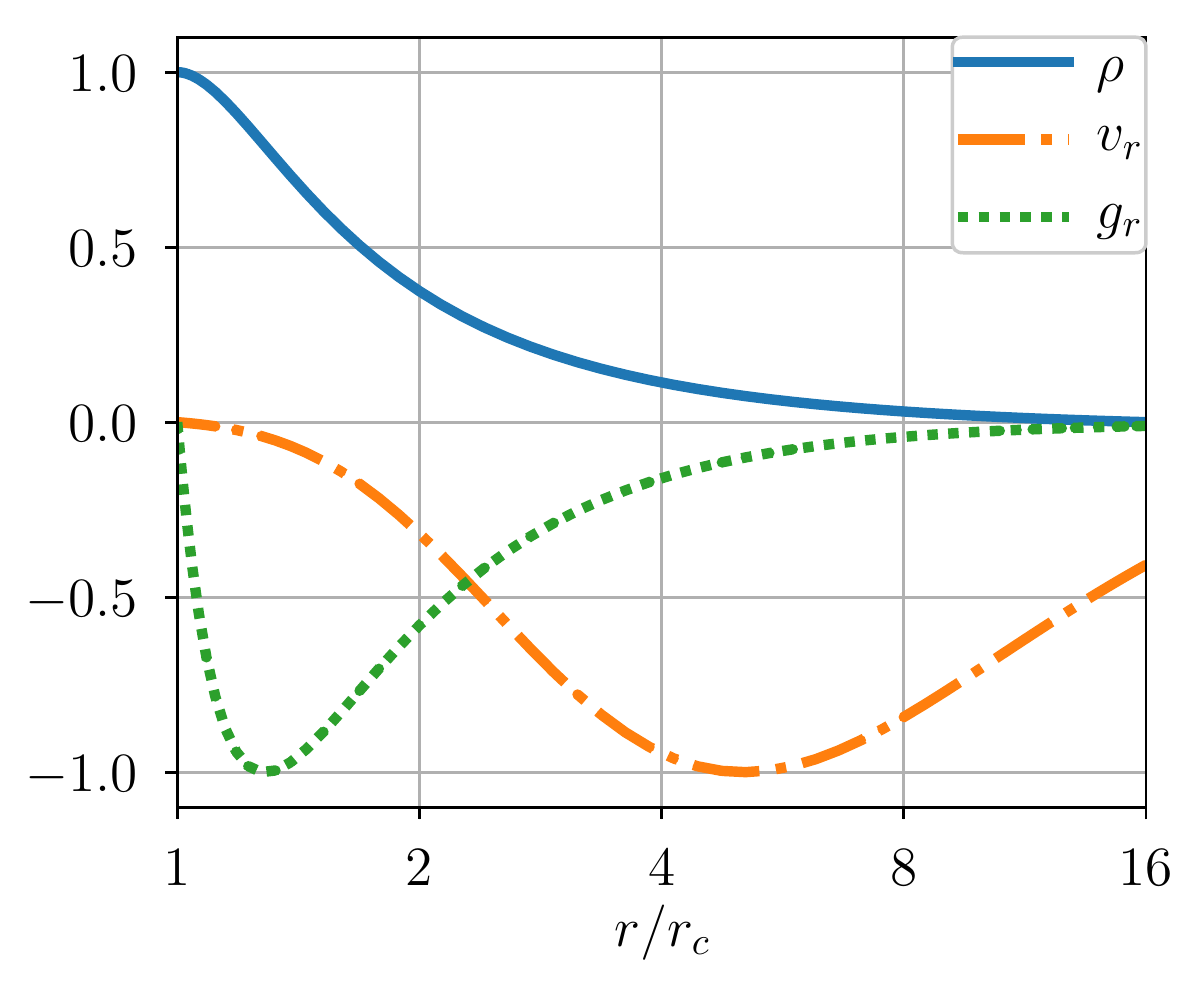}
\caption{Linearly unstable mode of an isothermal envelope of radius $\rO/r_c=16$, with a core mass $\rB/r_c=8$ and an gas mass $m_g = 0.46 m_c$, close to the critical value. The corresponding growth rate is $\omega r_c / c_s \approx 1.03\times 10^{-2}$. Each curve is normalized by its extremal value on the interval. \label{fig:eigenmode}}
\end{center}
\end{figure}

As long as $m_g < m_{\mathrm{critical}}$, the spectrum of \eqref{eqn:dtrho}-\eqref{eqn:dtgt} is purely imaginary, corresponding to gravito-acoustic oscilations of the envelope at quantized wavelengths. When $m_g \geq m_{\mathrm{critical}}$, the pair of eigenvalues corresponding to the largest-scale modes become real, leading to a contraction of the envelope as illustrated on Fig. \ref{fig:eigenmode}. 

\begin{figure}
\begin{center}
\includegraphics[width=1.0\columnwidth]{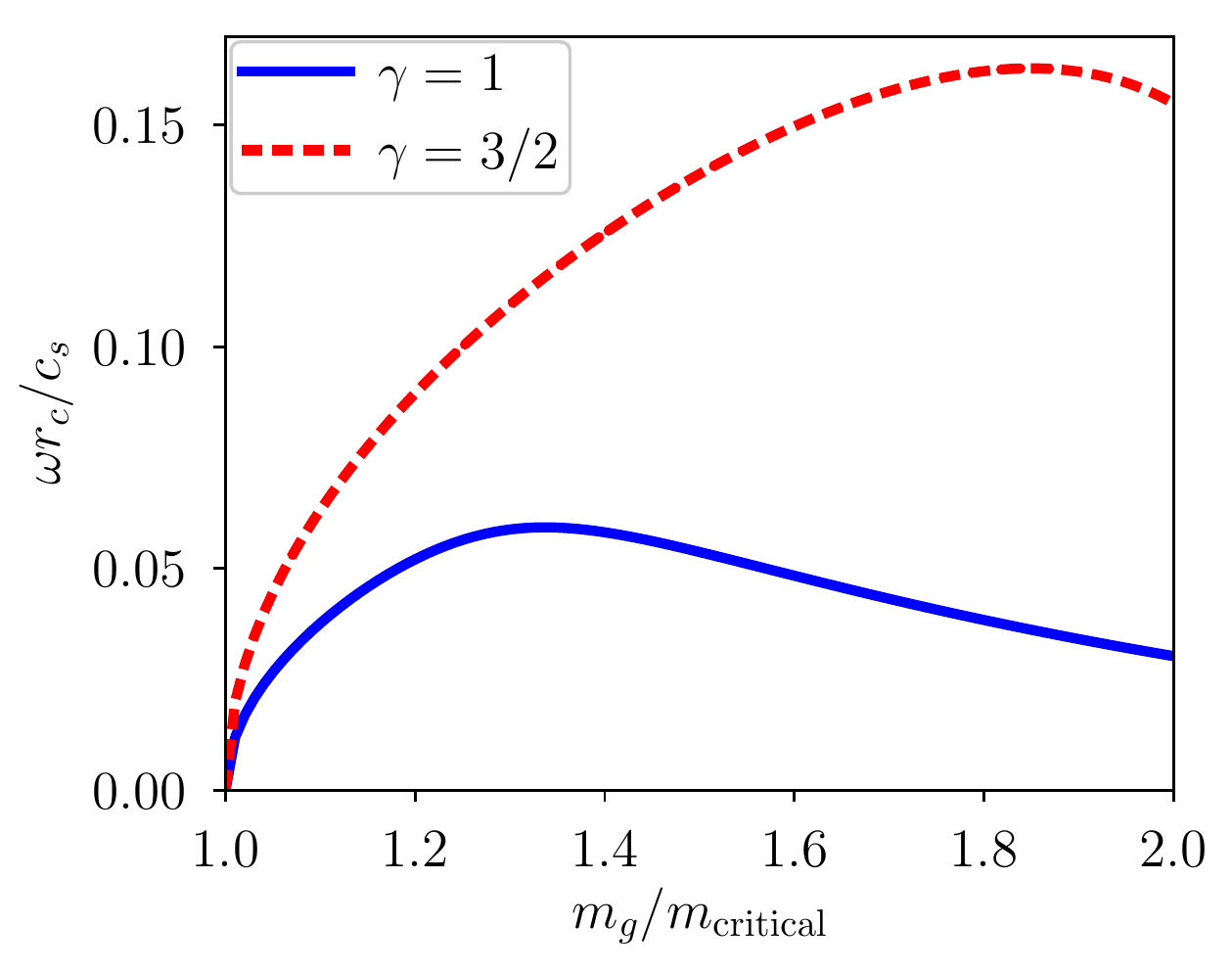}
\caption{Real eigenvalues from the system \eqref{eqn:dtrho}-\eqref{eqn:dtgt} as a function of the envelope mass $m_g$ and for different polytropic exponents $\gamma$ (see legend). \label{fig:growthrates}}
\end{center}
\end{figure}

Fig. \ref{fig:growthrates} shows the growth rates $\omega\in\mathbb{R}^+$ obtained when varying the polytropic exponent and the envelope mass. The growth rates are only a fraction of $c_s/r_c$, i.e. slow compared to the sound-crossing or free-fall time of the envelope. The case with $\gamma=3/2$ also features unstable modes, with even larger growth rates compared to the $\gamma=1$ case. In both cases, the growth rate has a maximum as a function of the envelope mass. 

We find linearly unstable modes for $\gamma=3/2$, whereas a homogeneous self-gravitating ball of gas should be linearly stable for $\gamma>4/3$. The larger growth rates found in the $\gamma=3/2$ case, as well as the presence of a maximal growth rate can be interpreted from a mean-field point of view. The equilibrium density profile $\bar{\rho}$ become steeper as $m_g$ increases or $\gamma$ decreases, such that the average Jeans length $\ell_{\mathrm{J}}$ eventually decreases over the scale $\rO$ of the unstable mode.


\bsp	
\label{lastpage}
\end{document}